\documentclass[12pt,preprint]{aastex}

\begin{document}

\title{ASTROPHYSICS OF YOUNG STAR BINARIES}

\author{{L. PRATO\altaffilmark{1,2}, T. P. GREENE\altaffilmark{2,3},
AND M. SIMON\altaffilmark{2,4}}}

\altaffiltext{1}{Department of Physics and Astronomy, UCLA,
Los Angeles, CA 90095-1562; lprato@astro.ucla.edu}
\altaffiltext{2}{Visiting astronomer at the Infrared Telescope
Facility, which is operated by the University of Hawaii under
contract from the National Aeronautics and Space Administration.}
\altaffiltext{3}{NASA/Ames Research Center, M.S. 245-6 Moffett Field,
CA 94035-1000}
\altaffiltext{4}{Department of Physics and Astronomy, SUNY,
Stony Brook, NY 11794-3800}

\begin{abstract}

This paper describes our study of the astrophysics of individual 
components in close pre$-$main-sequence binaries.  We
observed both stars in 17 systems, located in 4 nearby star forming
regions, using low-resolution (R$=$760), infrared spectroscopy and photometry.
For 29 components we detected photospheric absorption lines and
were able to determine spectral type, extinction, $K$-band
excess, and luminosity.  The other 5 objects displayed featureless or pure
emission line spectra.  In $\sim$50 \% of the systems, the
extinction and $K$-band excess of the primary stars dominate those of the
secondaries.  Masses and ages were determined for
these 29 objects by placing them on the H-R diagram, overlaid
with theoretical pre$-$main-sequence tracks.
Most of the binaries appear to be
coeval.  The ages span 5$\times$10$^5$ to 1$\times$10$^7$
years.  The derived masses range 
from the substellar, 0.06 $M_{\odot}$, to 2.5 $M_{\odot}$, and
the mass ratios from $M_2/M_1=0.04$ to 1.0.  Fourteen stars show evidence of
circumstellar disks.  The $K$-band excess is well correlated with the
$K-L$ color for stars with circumstellar material. 

\end{abstract}

\section{Introduction}
 
In the past decade, large numbers of young  
binary stars in dark cloud complexes such as   
Taurus and Ophiuchus have been identified
(e.g., Ghez et al. 1993; Leinert et al.  
1993; Reipurth \& Zinnecker 1993; Simon et al. 1995).  Indeed,
the frequency of these multiple systems is found to be at least
as great as that of the local field star population, and much
higher in the case of Taurus \citep{duq91}.  In order to
explore the astrophysics of the individual components in these
multiple systems, we initiated a program to determine their
properties: extinction, veiling, effective temperature ($T_{eff}$),
and luminosity ($L$).  Masses and ages are calculated from these
parameters.  Mass is the most fundamental
determining characteristic of a star; knowledge of masses  
and mass ratios in young star binaries provides insight into the  
binary formation mechanisms at work, and improves
our understanding of the initial mass function (IMF)
in these star forming clouds.  Ages are important because we wish  
to determine if the stars in relatively close binaries are  
coeval in order to understand what coevality, or lack thereof,
implies for models of binary star formation.  \citet{har94} found
about two thirds of a sample of 26 wide ($3''-45''$)
pre$-$main-sequence (PMS) to be coeval.  Studying a sample of closer
pairs, our approach is to choose a model of PMS evolution and test
the coevality of our sample on these tracks.

\citet{bra97} studied 14 young binary systems, observing at
visible light wavelengths.  They estimated masses and ages for 8
of the systems in their sample; mass ratios ($M_2/M_1$) of the components in
these binaries range from 0.5 to 0.8.  \citet{ken01} find a wider
range of mass ratios, $\sim$0.25$-$1, in their preliminary 
$HST$ STIS study of about
a dozen PMS binaries, with separations of $<$1$''$, in the Taurus star forming
region. Our approach was to study the astrophysics of
young binaries in infrared (IR) rather than visible light, providing
greater sensitivity to cool, red, low-mass companions and enabling
the measurement of mass ratios as small as
0.04.  In general, the seeing conditions are better in the     
near-IR in comparison to visible wavelengths, yielding
improved spatial resolution for observations
of close binary stars, in spite of the larger diffraction limit.
The extinction is also less at longer wavelengths.

Moderate resolution $K$-band spectroscopy is useful
for the characterization of young stars \citep{gre95, ali96, gre96, luh98}.
This spectral region encompasses a variety of atomic and     
molecular features, such as Na~I, Ca~I, and
the CO $\Delta v =$ 2 bandhead, which lend themselves well to
the identification of G through M type stars \citep{luh98, ali96}.
The Br $\gamma$ line of hydrogen serves
as a surrogate diagnostic for H$\alpha$ \citep{pra97, muz98}, 
and the shock
induced molecular hydrogen (H$_2$) lines across the $K$-band
trace outflows in the circumstellar environment.

We obtained spectra of systems in the Taurus, Lupus, Ophiuchus,
Corona Australis, and Aquila star forming regions (SFRs).  
We report here the results from the latter 4 regions,
in which both stars in 17 systems were observed; 2     
systems were observed twice.  The results of the Taurus
observations are in preparation (Prato et al. 2003).
In \S 2 we describe our spectroscopic and photometric observations
and data reduction.
Section 3 details our approach to estimating the stellar
properties for all objects with absorption
line spectra.  A discussion of the results is
provided in \S 4 and the conclusions are summarized in \S 5.
Color transformation equations and
comments on individual sources appear in the Appendices.

\section{Observations and Data Reduction}

\subsection{Observations}

\subsubsection{Spectroscopy}

The spectroscopic sample comprised 17 binary systems in the
Lupus, Ophiuchus, Corona Australis, and Aquila
regions with separations between
1$\farcs$2 and 7$\farcs$6.  These were selected on the basis of having
separations sufficiently large to be spatially 
resolvable by a long slit spectrometer
and brightnesses that enabled total integration times of
$<$1$-$2 hours in the near-IR.  About half of the sample is composed
of classic T Tauri stars (CTTs), i.e. systems with W(H$\alpha$)$\ga$10 \AA;
the others are weak-lined systems (WTTs) with W(H$\alpha$)$<$10 \AA.
Table 1 gives the object name in
column (1), an alternate name in column (2), right ascension and
declination in columns (3) and (4), the total $K$-band 
magnitude in column (5), the binary separation and position
angle in columns (6) and (7), the year of observation in column
(8), and the star forming region in column (9).
Unless otherwise noted,
the source of the $K$-band magnitude in column (5) was the photometry
described in the next section.
Separations and position angles were obtained from
Simon et al. (1995) and Reipurth \& Zinnecker (1993).
With the exception of AS 353, which lies in the galactic plane, all
of the SFRs studied in this project lie at galactic latitudes of
$|b|=10^{\circ}-20^{\circ}$.  Assuming the
field star density in these regions
is similar to the 5$\times10^{-5}$ stars per arcsecond squared found for
Ophiuchus \citep{gre92}, there is a 5 \% probability that one
of the systems in our sample is not a true binary.

Fifteen spectral type standard stars were
also observed.  This sample consisted of nearby,
main-sequence field stars with
spectral types of G2V to M8V, little or no extinction, and
approximately solar metallicity in most cases.  Standard
star properties appear in Table 2, which lists the object name in
column (1), the spectral type in column (2), and the $K$-band
magnitude in column (3).  With the exception of the one subgiant in
the sample, these objects and their properties were  
culled from an unpublished list compiled by J. Carr 
(private communication, 1994),  
derived primarily from \citet{kir91} and \citet{kee89}.  The subgiant,
HR 8784, a G8IV, was selected from
the Bright Star Catalogue \citep{hof82}.

We obtained all spectra at the University of Hawaii (UH) 2.2 m telescope
in 1996 July 3, 4, and 7 (UT) and 1997 May 28 and 29 (UT), with KSPEC,  
the former facility near-IR spectrometer \citep{hod94, hod96}.
KSPEC was equipped with a 1024$\times$1024 pixel  
HgCdTe HAWAII detector.  With the telescope in its f/31  
configuration, the plate scale was 0$\farcs$167/pixel.  
By means of a prism cross-disperser, the spectral range of
KSPEC spanned the $J$-, $H$- and $K$-bands simultaneously.  Only the     
$K$-band data is used in this work since it contains
the most useful variety of spectral  
features for characterizing extinguished, late-type young stars.
The slit width for all
observations was 0$\farcs$8, which yielded a spectral
resolution of about $R=\lambda/\Delta\lambda=760$; the slit length was $17''$.
In most cases the stars were observed individually.  In several
instances, however, the spectrometer slit was aligned with
both stars in a binary system and
their spectra observed simultaneously.  A standard beam-switch
of 6$''$ along the slit was used.  Exposure times varied
from $\sim$1 to 240 s per frame.  An external Argon
lamp was observed to calibrate the wavelength scale.
Dome flats and dark frames were also obtained.

All 1997 observations except those of WSB 28 (for want of a guide
star) were made with the facility tip-tilt adaptive optics  
system \citep{pic94, jim00}.  Tip-tilt accomplishes     
a first order wavefront correction to stellar images  
distorted by the Earth's atmosphere.  
The full width half maxima (FWHM) of the spectra for
the objects measured with tip-tilt  
on were consistently sub-arcsecond.  With tip-tilt  
off, the object point spread functions (PSFs) were $\sim$1$''$ wide.  

\subsubsection{Photometry}

JHK photometry for most of the objects was
obtained with QUIRC, the facility near-IR camera at the 
UH 2.2 m telescope, on 1996 July 1 (UT).  Conditions were  
photometric.  The QUIRC detector is a 1024$\times$1024 HgCdTe HAWAII
array \citep{hod96}.  In the f/31 configuration, the plate  
scale is 0$\farcs$061/pixel.  For each object
at each bandpass we took 4 exposures     
dithered by 7$''$ to place them at 4 different positions  
on the array.  Integration times varied from $\sim$1 to 45 s.      
Flat and dark fields were also obtained.  IRAS 16231$-$2427,
an A0 star, was observed at
several different airmasses during the night in order to provide
photometric calibration.  Several objects were observed for which spectroscopy
was not performed.  The facility tip-tilt system was used for most
observations.
    
Additional photometric measurements were made at the NASA IRTF 3m telescope  
in 1996 April 14 and 15 (UT) with NSFCam, the facility near-IR
camera \citep{ray93, shu94}.  The conditions were photometric. 
NSFCam is equipped with a 256$\times$256 InSb detector array;     
observations were made at $J$-, $H$-, $K$-, and $L$- bands.  The
plate scale
was 0$\farcs$056/pixel for DoAr 24E, and 0$\farcs$15/pixel for all
other objects.  Integration times varied from 0.1$-$15 s.
Integration times for flat and dark frames were 0.3$-$2.0 s.  HD 161903
(an A2 star) was
observed at airmasses similar to those of the target stars for
photometric calibration.  Objects in Lupus,     
as well as AS 205 in Ophiuchus, were not observed with either  
QUIRC or NSFCam.  For     
these systems, photometric data were taken from the literature
\citep{liu96, hug94}.

\subsection{Data Reduction}

\subsubsection{Spectroscopy}

All data reduction was done with IDL routines developed by one of us
\citep{pra98}.  A flat field was created by taking the difference
of 10 median filtered flat lamp exposures and 10 median
filtered dark frames.  Beam-switched pairs of stellar data
were differenced and divided by the final flat field.
These flat-fielded, differenced frames were inspected in order to     
eliminate data with low signal to noise, then
registered and averaged.  This final
average was cleaned of bad pixels by interpolation  
using the 4 neighboring pixels.  
   
For those cases in which one star was observed at a time, the
spectra were extracted by forming a model $K$-band PSF,
averaging more than 100 individual
columns in the cross-dispersion direction.
By fitting the model PSF to each of the individual 1024
columns across the $K$-band and minimizing the least     
squares residual, weighted by the pixels with the maximum counts, 
the spectrum at a particular wavelength (pixel column)
was extracted from the maximum value of the best-fit PSF.

If two stars with a small separation were observed
simultaneously along the slit, the 
resulting two-dimensional spectra may overlap.
In these cases, it was necessary to use a single star, observed at a  
similar airmass as the target binary, for creating a model point
spread function.  Since the binary    
separation and position angle were previously known (Table 1), the  
single star PSF was duplicated and the original and the copy
superimposed at the appropriate separation to mimic
the binary PSF for a particular system.  The extraction procedure was
analogous except that, in this case, Poisson weighting was used  
in fitting the model binary PSF to the observations.  
The binary flux ratio was determined from varying the PSF component
amplitudes and finding the best fit to the data.
Once the spectra were extracted,
any remaining spurious bad pixels were cleaned by interpolation.    

Internal reflections in the sapphire detector  
substrate of KSPEC gave rise to a variable frequency,
interference fringing pattern across the array \citep{hod96}.
To remove this pattern, the raw
spectra were filtered in the Fourier domain.
Differences in the equivalent widths as a result of the
filtering were on the order of a few percent or less and thus
significantly smaller than the uncertainties in the equivalent
width measurements.
    
The young star and spectral standard star spectra were divided by    
late B or early A type stellar spectra, taken at  
similar airmasses, to remove atmospheric    
and instrumental features.  The B and A star
spectra appear featureless in the $K$-band
except for broad  
Br $\gamma$ absorption lines.  In order to remove 
this hydrogen feature, a two-part process
was followed.  We
first interpolated over the Br $\gamma$ line in a G dwarf
spectral type standard star, and then inserted this
region of the G star spectrum into
the B or A star spectrum.  To restore the true
form of the spectrum, modified in the    
division by the early type telluric standard, it
is necessary to multiply the  
final divided spectra by a $K$-band black body curve of  
temperature corresponding to the spectral type of the telluric star.    
    
The wavelength scale was created by identifying the
column positions of argon lamp lines with known wavelengths and  
subsequently solving for the dispersion    
with a 2nd order polynomial.  The arc lamp column positions were
identified using one 4 s    
exposure taken during each observing run
and extracting an arc lamp spectrum from the same position
on the detector where the stellar data typically fell.
The sequence of reduced, wavelength calibrated
spectral type standard spectra appears
in Figure 1.  In Figure 2, the spectra of the young binary stars
are plotted.

\subsubsection{Photometry}

The QUIRC and NSFCam data were reduced in a similar manner.
Successive pairs of    
exposures of each object were differenced for sky subtraction,
flat-fielded, and cleaned of bad pixels, as for the spectroscopic data.     
For large separation systems, aperture photometry was obtained
for the components individually.
Otherwise, the photometry of the entire system was obtained and small    
apertures of 0$\farcs$2$-$0$\farcs$6 used on the central portions of  
each of the binary components to determine the    
binary flux ratio and hence the individual magnitudes.
As part of this process, airmass corrections, derived from the
photometric standards, were applied.

Table 3 lists all the photometry for the complete KSPEC sample  
as well as for several systems for which 
no spectroscopic data were obtained.  The first entry for each system
is the primary, indicated with ``A'', and the second is for the
secondary object, ``B'', as determined by \citet{rei93} and \citet{sim95}.
The uncertainties derive from the standard deviation of the mean of the
number of counts in an aperture for a set of four or five
background-subtracted exposures.
In order to present the photometry on a consistent basis, we
transformed from the QUIRC to the CIT systems.  Details of the
transformation appear in Appendix A.

Comparison of our JHK photometry
with that of \citet{gre92}, \citet{bar97}, and values from the
$2MASS$ database for objects common to our sample
usually yields results consistent to within 20 \% and, in many cases,
to within 5 \%.  For some objects, for example DoAr 24E, variations
of 20$-$50 \% were present between our photometry and that of
\citet{gre92} and \citet{bar97}.  However, agreement between our
results for DoAr 24E and those of $2MASS$ was better than $\sim$10 \%.
In general, our photometry best matched the $2MASS$ data;
for $\sim$80 \% of the objects or systems in common, agreement was
better than 10 \%.  Because T Tauri star magnitudes are known to vary
considerably, these results are not surprising.  The $J-H$ and $H-K$
{\it colors} of the stars in our sample are in good agreement, to within
1 $\sigma$, with colors based on the
photometry found in the literature \citep{gre92, bar97, luh99,
kor02}.

\section{Stellar Properties}

\subsection{Color-Color Diagram}

Figure 3 shows a color-color diagram
based on the $J$-, $H$-, and $K$-band photometry from Table 3;
all objects appearing in this Table are plotted
except for YLW 15A (which we did not detect at $J$; see Appendix B).
Colors for the dwarf and giant loci are from \citet{tok00}, transformed to
the CIT photometric system.  Figure 3 also shows the
CTTs locus from \citet{mey97}.
Combining the equations (5) and (6)
from Greene \& Meyer (1995) with the first equation in \S 2.2
of \citet{mey97} yields

$A_v$ $=$ 13.83 (J$-$H)$_{obs}$ $-$ 8.29 (H$-$K)$_{obs}$ $-$ 7.43.

\noindent

This derivation of $A_v$ was used as a guide only; a quantitative
treatment of the extinction appears in the next section.
Using both the $J-H$ and the $H-K$ colors, we estimate the $A_v$'s
of our sample, dereddening to the CTTs locus.  For the WTTs in
our sample, which, unextinguished,
presumably lie between the CTTs and the dwarf
star loci, this approach may underestimate the magnitude of the
extinction, depending on the stellar spectral type.

Three stars, the primaries in Elias 2-49, AS 353, and AS 205,
appear in an unexpected location, $>$1 $\sigma$
below and to the right of the CTTs locus in the color-color
diagram, at higher $H-K$ and lower $J-H$ values.
One of these, Elias 2-49 A, an A1V star, is not a member of our
spectroscopic sample so we will not discuss it here.  Of the
remaining two, AS 353 A is an unusual emission
line object.  In addition to Br$\gamma$, Na I, and CO bandhead
emission, the 2.059 $\mu$m line of He I is in emission in this object
(Figure 2).  Given that this is such an active source, the near-IR
colors may be abnormal as a result of bound-free and
free-free emission or scattering \citep{lad92}.  The
third object in this region, 
the AS 205 primary, is similarly a strong Br$\gamma$ emission line source.

\subsection{Effective Temperature, IR Excess, and $A_v$}

The observed spectra result from a combination of the
young star photosphere, excess IR radiation 
from circumstellar disk material, IR emission from
accretion processes, and interstellar extinction along 
the line of sight.  For strong emission line objects,
free-free and bound-free emission may also be present
and contribute to the spectra.  However, for absorption line
objects with determinable spectral types, bound-free and free-free
contributions to the continuum are probably negligible.  Thus, we do not
take these processes into account.  To estimate the spectral types,
IR excess, and extinction, we create
model spectra of the underlying young star photosphere, beginning with
an observed young star spectrum and varying the parameters of extinction,
amplitude of the $K$-band excess,
and, in some cases, the slope of this excess.
We use a multi-parameter $\chi^2$ fit to compare the
modified young star spectra
to spectral type template spectra.  In this way,
both the absorption lines and the continuum flux contribute
to the determination of the spectral type, $A_v$, and $K$-band excess.

Figure 1 shows our observed library of template spectral type
standard spectra.  Gaps in this sequence were filled by averaging
adjacent spectra: an M1.5V template was produced by averaging
the M0V and M3V spectra, and an M6 was created by averaging
the M5 and M7 spectra.  These averaged spectra are not shown in
Figure 1.  Table 4 lists the
wavelengths (column 1) of the spectral features for which we
calculated the equivalent widths of the young stars and the template
standard stars in our sample.
Column (2) gives the line width used for this calculation.  The
species is listed in column (3), the details of the transition
\citep{kle86, tok00}
in column (4), and comments in column (5).  The equivalent widths
measured on the basis of the parameters presented in Table 4 appear
in Tables 5 and 6 for the standard template stars and the young
stars, respectively.

To determine the suitability of main-sequence spectral type
standards for the characterization of our young star sample,
we followed the approach of \citet{gre95}, illustrated in
their Figure 4, and compared the positions of the standard and
target stars on a plot of the Na I$+$Ca I {\it vs} CO(4$-$2)$+$CO(2$-$0)
equivalent widths.
The dwarf locus in our Figure 4 was determined from fits to the equivalent
width data found in Table 5, excluding the
G8IV, M7V, and M8V standards because these objects departed from a linear
fit.  The linear region included data from
the G2.5 through the M5 standard stars.  The G8IV is a sub-giant and
so we did not expect it to conform to a linear fit to the dwarf
standard stars.  The 1 $\sigma$ scatter in the dwarf
star equivalent width values around the best fit line
plotted in Figure 4 was $\sim$1 \AA, which probably reflects some
intrinsic scatter in the heterogenous sample of
standards.  Hence, we characterize the
uncertainty in the equivalent widths appearing in Tables 5 and 6
as 1 \AA.  Because the equivalent widths of many lines
were detected only at the
1 $\sigma$ level, spectral type determination using line ratios
was not possible \citep{pra98}.

The giant star locus was
defined using data from \citet{wall97}.  Digital spectra from
the companion CDROM to that paper were smoothed with a Gaussian
filter to reduce the spectral resolution from the original
R $\sim$ 5000 to that of our KSPEC sample, R $\sim$ 800.
The resulting equivalent widths were then measured as for the dwarfs and used
to determine the giant locus.  In Figure 4, the
distribution of the young binary stars (squares) around the dwarf locus
indicates that these standards provided adequate
templates for the characterization of spectral types.

To identify the young star spectral types, a procedure was used which
iterated through a range of values for $A_v$    
and a constant (zero slope) $K$-band excess, creating
several thousand modifications of each young star photospheric spectrum
according to

f$_{photosphere}(\lambda)=(f_{obs}(\lambda)e^{\tau_{\lambda}}-k)\times c$  
  
\noindent where f$_{obs}$($\lambda$) is the observed spectrum of young star,
$\tau_{\lambda}=0.92A_{\lambda}$,
$k$ is the contribution to the $K$-band
in excess of the stellar photosphere, and c is a scaling
constant which takes into account differences in the distances and radii
between the spectral template standard spectrum and the model
young star spectrum.  
To define $A_{\lambda}$, we used a linear fit to
the van de Hulst no. 15 \citep{joh68}
curve between 2 and 2.5 $\mu$m and obtained a $\lambda^{1.6}$  
dependence, consistent with \citet{rie85}.
Therefore, we define
  
$A_{\lambda}=(0.55/\lambda(\mu m))^{1.6}A_V$.  

To first order, we approximated the $K$-band excess
as a constant contribution across the band.  
We determined the values of $A_v$ and the $K$-band
excess by the method of least squares.  For a given value of the
constant, $c$, the $A_v$ and $k$ parameters were varied
to minimize the residuals between the YSO spectrum and a template
spectrum (Figure 1).  A wide range of
values for $c$ and most of the spectral type standard stars were always tested
in order to determine the best match.  Visual inspection of the ten best fits
of each modified young star spectrum to each standard star aided the
final choice of best fit.  In general, the parameters
of these ten were similar and often visually indistinguishable.

For a few objects, the
young star photospheric spectra could be well-fit to either an early
type standard, by accounting for a large $A_v$ with little veiling, or
to a late type standard, if the value of the $A_v$
was small but the veiling greater.  In
such cases, estimates of $A_v$ from the color-color diagram
(\S 3.1, Figure 3) were used as a guide to the most accurate fit.
These objects are discussed in Appendix B.

We estimated the 1 $\sigma$ uncertainties by determining
at what values of $A_v$ and $k$ the normalized $\chi^2=3.2$
\citep{wal96}.  Figure 5 is a contour map of $\chi^2$
as a function of $A_v$ and $k$, for a particular
value of the constant $c$ for which the $\chi^2$ was minimized.
The figure shows the fit of the modified spectrum of
ROX 15 B to the
M3 standard star spectrum.  Standard stars of spectral types K7 to M5 were
also used, but the fit to the M3 clearly produced the best apparent fit and
the minimum $\chi^2$.

For some objects, a constant $K$-band excess did not produce as
precise a fit to the spectral type standards as a wavelength dependent
$K$-band excess.  For these cases, we modelled the excess as a linear function
of wavelength where

f$_{photosphere}(\lambda)=(f_{obs}(\lambda)e^{\tau_{\lambda}}-k(\lambda))\times c$

\noindent and
    
$k(\lambda)=m\lambda+b$,    
    
\noindent is the non-uniform $K$-band continuum excess which
gives the best fit; m and b are positive constants.  This applied to 5 stars,
WSB 4 B, WSB 19 A, WSB 28 B, and S CrA A and B.  The latter two
reveal the largest $K$-band excesses observed in our sample.  WSB 28 B is
apparently substellar (\S 4.1); WSB 4 B and
WSB 19 A are unexceptional.  At
wavelengths longer than $\sim$3 $\mu$m, the IR excess is often well fit by a 
power law of index $n=0.75$ \citep{shu87}.  However, the spectral energy
distribution varies from star to star between 2 and 3 $\mu$m
(e.g. Strom et al. 1989), probably as a function of circumstellar
dust temperature and the level of accretion activity.  Therefore it is not
surprising that we find positively sloped excesses in some cases.

Once the best values for spectral type, $A_v$, and $k$
were found, the ratio of the magnitude  
of the $K$-band excess to the photospheric flux of the    
star at 2.22 $\mu$m, $r_k$ = F$_{K_{ex}}$/F$_{K_*}$ = $k$/F$_{K_*}$,
was calculated.  The effective YSO temperature, $T_{eff}$, was derived
from the spectral type $-$ $T_{eff}$ scale illustrated in
Figure 5 of \citet{luh00}.  Table 7 lists these derived sample
properties with spectral type in column (2), $T_{eff}$ in column (3),
$A_v$ in column (4), and $r_k$ in column (5).
For a few of the young binary components, it was not possible to
determine a good fit to spectral type
standards with any combination of parameters (see Appendix B).
In these cases the uncertainties in spectral type were
three spectral subclasses.  Otherwise, typical uncertainties in 
spectral type were one to two spectral subclasses.

Late M star spectra are dominated by the CO bandheads,
which appear at wavelengths
longer than $\sim$2.3 $\mu$m and cover about
a third of the spectral region which we use for the spectral
type determination described above.  These features are particularly
sensitive to surface gravities \citep{kle86}. However, because
the young star targets are associated primarily with the
dwarf locus in Figure 4, we are confident that low surface
gravities are not effecting our results significantly.
This is probably because, by the time a star is recognizable
as such, most of the contraction to the main-sequence has taken
place for the systems explored here \citep{bar98}.

Paradoxically, \citet{gre96} and \citet{mam02} find evidence for
lower surface gravities in the $\rho$
Ophiuchus SFR and the Sco-Cen OB complex,
respectively.  Although we use the same criteria here as \citet{gre96},
their dwarf luminosity class locus (their Figure 4)
has a steeper slope than ours
and was derived from a smaller sample of standard main-sequence stars.
\citet{mam02} use completely different criteria for the determination
of luminosity class, and, hence, surface gravities.  Detailed 
comparisons of identical diagnostics for large samples of dwarf
and giant standard stars and young stars from a variety of star
forming regions with well determined ages will be necessary to
understand these discrepancies.

\subsection{Luminosity}

Using $T_{eff}$, $A_v$, and $r_k$,    
together with the $J$-, $H$- and $K$-band
photometry, the luminosity was calculated.
The procedure followed corrects the observed
$J$-, $H$- and $K$-band magnitudes for the derived value of $A_v$, and the  
$K$-band magnitude for the calculated $r_k$.  We did not correct for
excess continuum emission in the $J$- and $H$-bands.  At $J$,
we assume that the excess
is effectively negligible, and at $H$, it is
typically less than half of $r_k$ (Greene \& Meyer 1995).
A black body curve,    
determined by the stellar $T_{eff}$, was scaled to   
match the dereddened photometry.  
The integral under this curve provided
an estimate of the luminosity, given in column (7) of Table 7. 

The largest internal source of uncertainty in the final value
of the luminosity derives from the distances to the
SFRs under study.  The values used appear in column (6) of
Table 7; the references for these values appear in column (8).
Unless the uncertainty was given explicitely in the reference,
we used $\pm$20 pc.  The uncertainties
resulting from errors in the photometry, $T_{eff}$, $r_k$, and $A_v$
were smaller.  Another source of error in the late M star luminosities
may be attributable to the fact that the 
photospheres of cooler stars deviate substantially from a Planck function, 
depending on the metallicity \citep{all97}.  Thus, we could
be overestimating the luminosities of the latest M stars.
As a check on our estimates for the luminosities, we also
followed the procedure described in \citet{gre95}
for calculating luminosities from the $J$-band magnitudes
and bolometric corrections.  Our approach generally yielded larger
estimates for the luminosities by a factor of 10$-$30 \%, however,
there was no trend for this discrepancy to be larger in late-type
objects.  The presence of unresolved companions is another source of
uncertainty in the determination of luminosity; a few objects in our
sample which are likely to be triples are discussed in more detail
in Appendix B.

\section{Discussion}

\subsection{Ages, Masses, and Mass Ratios}

With values for $T_{eff}$ 
and luminosity from Table 7, we plot the objects in our sample on
the PMS evolutionary tracks overlaid on the H-R diagram to estimate
their ages and masses (Table 8).  In 
the upper panels of Figure 6 we show the
young star binary components plotted on the H-R diagram
with the PMS tracks of \citet{pal99} (left) and \citet{bar98} (right).
The objects cover a wide range of ages and masses, from
$\sim$1$\times$10$^5$ to 1$\times$10$^7$ years and 0.06$-$2.5 $M_{\odot}$.
They fall more completely on the younger, higher mass tracks
of \citet{pal99}, thus, in Figure 7, we show individual H-R diagrams for
each binary pair plotted on these tracks.

Following the recommendations of \citet{sta01},
we derived $T_{eff}$ and luminosity
for a sample of 14 single, main-sequence standard stars. This sample
was culled from the list of spectral type standard stars which
appears in Table 2 of \citet{pra02}.  The spectral types of these
objects ranged from G0 through M9.  In most cases,
their metallicities are
solar and they are confirmed radial velocity singles.
We determined their spectral types from the high-resolution
$H$-band spectroscopy, shown in Figure 1 of
Prato et al. (2002).  The values of $T_{eff}$ for the main-sequence
dwarf temperature scale
were obtained from \citet{tok00} for the G and K type standards
and from \citet{leg00} and \citet{leg02} for the M type
stars.  Uncertainties in the spectral types were
typically 1 subclass.  The luminosities were
calculated in the same way as described in \S 3.3.  We used distances
derived from the {\it Hipparcos} parallax, as given in the 
Centre de Donn\'ees astronomiques de Strasbourg (SIMBAD)
database, and $J$-, $H$-, and $K$-band magnitudes from
\citet{leg92}, \citet{gez93}, and the colors given in \citet{tok00} 
in conjunction with apparent $V$-band magnitudes listed in SIMBAD.
Uncertainties in the
luminosity reflect only uncertainties in $T_{eff}$ and, hence, in
the spectral subclass, which dominated the errors in the luminosity
calculation.

In the lower panels of Figure 6 we show the positions of 
the 14 single, main-sequence standards described above on the H-R diagram.
Isochrones only as old as 1$\times$10$^8$ years are plotted.
All 14 stars fall within 1 $\sigma$ of the
3$\times$10$^7$ year isochrone on both
the \citet{pal99} and \citet{bar98} tracks.
This is unexpected, given that the typical ages of this main-sequence
sample are thought to be 1$-$5$\times$10$^9$ years.
In general, the lower mass main-sequence stars appear to be slightly younger.
In \S 3.3, comparison with an alternative approach to luminosity
estimates did not reveal any mass bias in our estimates of the luminosities,
although in general our approach yields larger luminosities than the
technique of estimating $L$ from the bolometric correction and the
$J$-band magnitude \citep{gre95}.  A systematic overestimate of
the stellar luminosities may therefore be responsible for the young ages
found for the main-sequence sample on both the tracks of \citet{bar98}
and those of \citet{pal99}.  For the stars in our PMS
spectroscopic sample, this would imply underestimated ages.  Alternatively,
the spectral type-effective temperature conversions used for the
main-sequence standards could have led to temperatures which are
systematically too cool by 200$-$300$^{\circ}$ K.

Out of the 17 young binaries
observed, it was possible to derive masses and ages
for both components in 13 binaries, for one component in 3 binaries, and
for neither component in the VV CrA system.  Thus, we have derived
masses and ages
for 29 out of the 34 stars in the sample.  
A histogram of $q=M_2/M_1$ for 13 systems appears in Figure 8.
Although the sample is statistically small
and necessarily incomplete, most of the mass ratios are distributed
between $q=0.05$ and $q=0.50$; no concentration towards $q=1$ is
seen, in contrast to the $q=1$ trend 
in the observed mass ratio distribution of 
young star spectroscopic binaries detected in visible light,
illustrated in Figure 4 of \citet{pra02},
probably the result of selection effects.

Figure 9 shows $q$ as a function of binary separation.  For the
smaller separation systems, the mass ratios range from 0.04 to 1.0.
However, for systems with separations greater than $\sim$3$''$, the
mass ratios are all $<$0.3.  Among this group of wide separation
and small mass ratio systems is the WSB 28 binary, which appears to
harbor a brown dwarf secondary with a mass of $\sim$0.06 $M_{\odot}$.
At the distance of 160 pc to Ophiuchus, the WSB 28 binary has a separation
of $\sim$800 AU.  This is consistent with the conclusions of
Gizis et al. (2001) that no brown dwarf companion desert is observed at
separations of 1000 AU and greater in the field star population.

Of the 13 systems with estimated ages for both components, 11 appear
to be coeval within the 1 $\sigma$ uncertainties.
The lack of coevality in 2 of the systems, SR 21 and WSB 71,
may be the outcome
of faulty characterization of the component stars, the
result of false binaries resulting from chance
alignment, or simply statistical variations in the observed
properties used to determine age.
We can calculate the probability of a spurious binary
in our sample in order to test the chance alignment
hypothesis.  With the exception
of Aquila, which lies in the galactic plane, all
of the SFRs studied in this project lie at galactic latitudes of
$|b|=10^{\circ}-20^{\circ}$.  Assuming the field star density in these regions
is similar to the 5$\times10^{-5}$ stars per arcsecond squared found for
Ophiuchus \citep{gre92}, there is a $<$5 \% probability that one
of the systems in our sample is not a true binary.  The apparently
non-coeval systems have two of the smallest measured mass ratios
in the sample; such low mass ratio
systems, if true binaries, provide good tests of the high and low-mass
regimes of the tracks.  Common proper motion studies of these
objects would confirm or reject the premise of true binarity,
improved measurements of the T$_{eff}$'s and, in particular, of  the
luminosities, as $\sigma$($L$) dominates the uncertainties,
would reveal discrepancies in our masses and ages, and high
resolution imaging would probe for higher order multiplicity.
It is also possible, but more difficult to test, that in these relatively
wide separation, non-coeval systems, environmental differences have
played a role in their uneven development.

\subsection{Component $A_v$'s}

The determinations of $A_v$ described in \S 3.2 and shown in
Table 7 were larger than the values found from dereddening the
objects to the CTTs locus in the color-color diagram (\S 3.1).
This systematic difference may be attributable to distinct origins
for the extinction measured by the two techniques.  Where the
color-color diagram measures the $A_v$ to the scattering surface in
a system, the spectroscopic approach measures the $A_v$ to the stellar
photosphere.  Because the photosphere is typically more obscured,
the spectroscopically measured $A_v$ is larger.

Inspection of the individual $A_v$'s measured for both stars in
the 13 characterized systems reveals that for 6 binaries,
HBC 248, WSB 28, DoAr 26, ROX 15, WSB 71, and HBC 679, the
component $A_v$'s do not agree within the uncertainties (Table 7).
In Figure 10 we plot the primary {\it versus} secondary star
$A_v$.  The primary star $A_v$'s are
usually larger when there is a difference in the component
$A_v$'s.  Most of the systems with different values of extinction
for each component have mass ratios of $q<0.2$.
Large differences in component masses suggest that
a higher mass primary star is associated with a deeper potential,
and therefore with a larger amount of circumstellar material
\citep{bat00, pra01, whi01}.  This is consistent with the association
of higher $A_v$'s with the primary stars.  \citet{whi01} did not
observe this result in their sample of objects with $A_v<5$ in the
Taurus star forming region.  They found that the component
$A_v$'s tended to be similar, consistent with the results of
\citet{lad99}, which suggest a homogeneous extinction.  Because
our observations measure the extinction to the stellar
photospheres, our results may reflect
differences in the amount and distribution of circumstellar material.

\subsection{Circumstellar Disks}

\subsubsection{General Sample Characteristics}

Dereddened $K-L$ colors provide excellent diagnostics for
the presence of circumstellar disks \citep{edw93, pra97};
\citet{ken95} find a typical $K-L$ variability
for young stars in Taurus of only $\sim$0.1 mag.  \citet{edw93}
define the presence of a circumstellar disk by a $K-L$ color in
excess of 0.2$-$0.4 mag.
In column (3) of Table 9 we list all the $K-L$ values available for the
spectroscopic sample, compiled from this work and from the 
literature (see Table 3).  These $K-L$ colors range from
$\sim$0 to 3.63.
In some cases only the unresolved, systemic color
was available; we include these for completeness.  Before
calculating the colors, the values for $A_v$ given in Table 7 were
used to deredden all of the $K$- and $L$-band magnitudes.
In column (2) of Table 9 the values for the 
Br$\gamma$ emission line fluxes
appear.  The conversion from equivalent width to line flux was
performed using the dereddened $K$-band magnitudes.

In some cases
the 1 $\sigma$ uncertainties in the $K-L$ colors are large
enough that the presence of a disk
is difficult to judge.  For these objects, comparison with the
expected photospheric colors is
helpful.  Results for individual objects are discussed in Appendix B.
Several stars with photospheric $K-L$ colors
and small $K$-band excesses manifest small hydrogen emission lines.  These 
may be chromospheric in origin \citep{wal94}
and are noted in column (4) of Table 9.
In the four cases for which only unresolved colors are available,
they are indisputably photospheric, indicating that neither
component has $K-L>0.2$.

Figure 11 shows the primary {\it versus} secondary $K-L$ colors.
No obvious correlation is present, in contrast to the results
of \citet{whi01}, however, our sample is small.  A plot of
primary {\it versus} secondary $K$-band excess, $r_k$, appears in Figure 12.
In this case, a trend towards primaries with higher $r_k$'s than
the secondaries is clear.  A natural gap between low and high 
$K$-band excess occurs at $r_k\sim0.3$.  Below this cutoff, the
excess appears negligible within the uncertainties.  
The trend towards higher values of primary star $A_v$ and $r_k$ implies
the presence of more circumstellar material around the
relatively more massive
stars.  This follows if infall onto the disks at the time of formation
originated in low-angular momentum flows from a spheroidal
circumbinary infall \citep{bat00}.

In Figure 13 we plot $r_k$ as a function of the $K-L$ color for
19 individual objects.  Almost every source with a large $r_k$ also
has a large $K-L$.  Because the primary star $r_k$ values dominate those of
the secondaries (Figure 12), a predominance of primary star colors in
Figure 11 is also expected.  The absence of such a result probably
derives from the small sample size for which both primary and
secondary dereddened $K-L$ colors are known.  In Figure 13,
objects with $r_k<0.3$ and $K-L<0.4$ mag
are associated with little or no circumstellar material
\citep{edw93}; {\it all objects with $K-L<0.4$ have $r_k<0.3$.}
With the exception of WSB 71, stars with $r_k>0.3$ and
$K-L>0.4$ mag display Br$\gamma$ line emission, indicating
that these objects have active, accreting circumstellar disks.  

Hydrogen emission lines vary greatly on short timescales \citep{gra92, pra97}
and do not always provide reliable disk diagnostics as they
can give not only false positive disk
detections, when in fact the emission is chromospheric in origin
\citep{mar98}, but also false negative results, when accretion
is in quiesence.
Compare, for example, the Br$\gamma$ spectra of the AS 205 and VV CrA
components shown in Figure 2 of \citet{pra97} and in Figure 2 of this
paper.  Statistically, objects 
with large $K-L$ colors often have large emission line fluxes.
No objects with small $K-L$ colors are observed to have large Br$\gamma$
line fluxes, although a couple of these objects do have moderate line
fluxes, probably attributable to chromospheric activity.

Using all available disk diagnostics,
$r_k$ values, $K-L$ colors, Br$\gamma$ emission lines, and, in one case, a
$K-12\mu$m color, we find that 5 binaries in our sample
are so-called mixed systems, in which
only one component is apparently associated with circumstellar material.
This is similar to the fraction of mixed systems found by \citet{pra01}
based on hydrogen emission lines after
applying corrections for systems with marginal CTT H$\alpha$ equivalent
widths and $K-L$ colors in excess of photospheric. Of
the 5 mixed systems, the two which appear to
be non-coeval, SR 21 and H$\alpha$71,
have separations of $\ga$5$''$, but the other three,
WSB 4, WSB 19, and DoAr 26, have separations of $\sim$2$-$3$''$
(320$-$480 AU) and appear coeval.  These three systems are
puzzling; it is not obvious why stars in binaries of a few hundred AU
separation which presumably formed simultaneously
should manifest diverse evolutionary states.  DoAr 24 E,
YLW 15A, and AS 353 are yet more extreme examples of this;
for these systems it was not even possible to determine the
properties of one component.  The geometry of these systems and the
orientation of the circumstellar disks may play a role in the
discrepant appearances of these binary components.

\subsubsection{WSB 4}

The 2$\farcs$8 binary, WSB 4, displays unusual properties in the
secondary star.  Some confusion has propagated in the literature regarding
the position angle (PA) of this pair.  \citet{zin92} show $J$-, $H$-, $K$-,
and $L$-band images of WSB 4 (their Figure 1), however the
orientation is not specified.  Their Figure caption indicates that the
northwest component is the optically brighter star, however, the figure
itself implies a PA of $\sim$40$^{\circ}$.  \citet{mey93} give
PA $=$ 310$^{\circ}$, inconsistent with the coordinates provided for
the two components in their Table 3.  \citet{rei93} also list
PA $=$ 310$^{\circ}$, but label the components in their Figure 1
such that the PA $\sim$ 130$^{\circ}$, consistent with Table 3 of
\citet{mey93}.  \citet{kor02} shows the same result.  B. Reipurth
(2002, private communication) verified the PA of 130$^{\circ}$ based on
the $R$-band image shown in \citet{rei93}.

Our photometry, in conjunction with the discrepancies in the literature,
initially compounded the confusion because we observed that for all three
bands, $J$, $H$, and $K$, the southeastern component was brighter
(Table 3).  \citet{mey93} and \citet{kor02} both show that at
wavelengths longer than $\sim$2 $\mu$m the southeastern component is
the brightest; it is also apparently significantly variable at shorter
wavelengths.  Because of the known variability in this system, the
$K$ and $L$ magnitudes used to calculate the
$K-L$ colors reported for the primary and secondary in Table 9 were
taken from \citet{mey93}.  In Figures 11 and 13, the much larger,
dereddened $K-L$ color of WSB 4 B, as compared to A, stands out.
WSB 4 B manifests only modest veiling, as measured by $r_k$.
We speculate that this object might be in the process of forming
an inner gap in the disk, with enough warm dust at the outer
radius of this gap to give rise to the very red color \citep{ken96}.
We observe a weak Br$\gamma$ emission line in WSB 4 B and \citet{mey93}
report an H$\alpha$ equivalent width of $\sim$20 \AA, indicating that some
accretion continues to take place.

\section{Summary}

We conducted a low-resolution (R$\sim$800)
$K$-band spectroscopic study of 17 angularly resolved young star binaries
in the Lupus, Ophiuchus, Corona Australis, and Aquila SFRs.  The spectra
reveal a variety of atomic and molecular features.
For 13 systems, we identified the
spectral types, extinctions, and $K$-band excesses of both
components, based on modelling of their photospheres 
and comparison with spectral type standards.  These standards, 
also observed as part of this study, range from G2.5V to M8V.
For 3 young binaries, it was only possible to identify
the spectral type of one component.  In the VV CrA system, neither
component displayed  absorption lines and thus their properties
could not be identified.  For those cases in which data from the
literature was available for comparision,
there was no obvious, systematic difference between the
visible light (unresolved spatially) spectral type and our primary
star spectral types.

Effective temperatures derived from spectral types and combined
with near-IR photometry, usually
taken at or close to the same time as most of the
spectroscopic observations, yielded 
luminosities.  All objects with assigned temperatures and luminosities
were plotted on the H-R diagram overlaid with PMS evolutionary tracks.
A sample of single, main-sequence stars, whose luminosities were
calculated in the same way as for the young stars, show anomalously
high luminosities. Thus, we suspect that our approach to the 
measurements of the luminosity produces an overestimate by $\la$50 \%.
A spectral type-temperature
conversion which causes main-sequence objects to appear to be too
cool would also cause the luminosities to appear larger than
expected for $1-5\times10^9$ year old stars.
Unresolved binary companions also can produce an
overestimate in the luminosity.  Koresko (2002) finds that several
binaries in out sample appear to have sub-arcsecond companions
(Appendix B).

The majority of the PMS systems are coeval on the tracks
of \citet{pal99}.  The two that are not coeval are among the wider binaries
in the sample, SR 21 and WSB 71.
The mass ratios for the 13 binaries with identified properties
range from 0.04 to 1.  None of the
mass ratios, in systems with separations $>$3$''$, are greater than 0.3.
The masses for all the stars in the sample which were assigned spectral
types span 0.06$-$2.5 M$_{\odot}$.

We examined the space density of field stars in the regions studied,
with the exception of Aquila, and
found that the typical probability for an interloper in one of the
systems observed is $<$5 \%.  The probability of a second
interloper is negligible.  Thus, the chance projection of a
background star probably does not explain the
non-coeval sources or the systems with very disparate
components, such as DoAr 24E and AS 353.  It is possible that 
these apparent component differences are attributable to
environmental inhomogeneities and
complex geometries of circumstellar material
in young star binaries.  This is supported by the
observation that many of the systems manifest differences in
the $A_v$ of the 2 components (Table 7); in these cases, it is
usually the primary star with the larger $A_v$ (Figure 10).

Uncertainties in the distances to both the SFRs and to the 
individual objects within these regions comprise an important
source of uncertainty in the calculation of the stellar
luminosities which could effect our final results.  
In combination with statistical
uncertainties in other observed young star properties, this
might account for the 2 non-coeval pairs.  In general,
our approach to the calculation of the luminosity
seems to produce systematic overestimates, which results in
underestimates of the stellar ages.

A least half of the objects in our sample appear to have circumstellar disks.
We used Br$\gamma$ emission lines, $K-L$ colors, and $r_k$ values
as diagnostics of circumstellar material.
We detected no evidence for strong dominance of primaries in a plot of
secondary {\it versus} primary dereddened
$K-L$ colors (Figure 11), in contrast to the
results of \citet{whi01}.  However, the sample for which dereddened
$K-L$ colors are known for both components is small.
We do see a trend towards larger
values of $r_k$ in primaries (Figure 12); the veiling as measured
by the $K$-band excess, $r_k$, may provide a better measure of circumstellar
material close to the central star than the $K-L$ color.
In general, the amount of circumstellar material around the primary
is expected to exceed that around the secondary star on the basis of
the stronger primary star potential.  Figure 13 depicts a clear
correlation between large values of $r_k$ as a function of $K-L$
color, implying that, for stars which have high levels of accretion,
and thus a large $r_k$, the inner disk is optically thick,
as measured by the large $K-L$ color.  One
exception to this pattern is WSB 4 B; a circumstellar disk
with a growing inner gap, producing a large $K-L$ color in the absence of
strong accretion, may explain this.

Five of the systems analyzed are so-called ``mixed'' pairs, in which
only one component appears to have circumstellar material
\citep{pra97, pra01, ken01}.  We used a
Venn diagram analysis similar to that of \citet{ken01} to 
compare the various disk diagnostics in the sample stars.
Not all of our objects have been observed in all three diagnostics,
either because the data used is incomplete, or because pure emission line
spectra did not permit the calculation of $r_k$.
However, using the existing data, we conclude that at least 17 of the objects
in the sample are CTTs and no more than 17 are WTTs, some of which may
reveal mid-IR excesses upon further observation (e.g., WSB 19 B).
Two of the mixed pairs, WSB 71 and SR 21, are the only non-coeval
binaries in the sample, and in
addition have two of the largest separations.
We speculate that the unmatched properties of the stars in mixed systems
were determined by crucial differences in the
local circumstellar environments.

YLW 15A (Figure 14) is remarkable in that it is obscured by a large
$A_v$ and is heavily veiled yet displays detectable 
absorption lines.  This result gives us confidence that with
low-resolution spectroscopy, R$\sim$800, we can recover
photospheric spectra veiled by as much as $r_k=2$.  However,
higher resolution spectroscopy, $R\ga3000$, is far more
sensitive to small absorption and emission lines \citep{gre02}.
This sensitivity increases with resolution.
Such detailed spectroscopy can help to break the
degeneracy found for some objects in our approach to estimating
their properties simultaneously (\S 3.2) and reveal the
underlying properties of more heavily veiled and obscured sources.

\bigskip
\section{Acknowledgements}

We thank Chris Stewart and John Dvorak of the UH 2.2 m telescope,
and Dave Griep and Charlie Kaminski of the NASA IRTF for their
technical support.  Tracy Beck reduced the photometric data taken
at the IRTF, and Saeid Zoonematkermani provided several IDL
procedures used in the data reduction and helped to concatenate
the panels in Figure 6 and 7; we are grateful for their
assistance.  Discussions with John Carr, Suzan Edwards, George Herbig,
Deane Peterson, Charlie Telesco, Alycia Weinberger, and Ralph Wijers
enriched this paper.  We thank Russel White for providing data in
advance of publication and are indebted to a meticulous, anonymous
referee who provided constructive comments which have improved this
work.  L. P. acknowledges generous
support from Ian McLean during the preparation of this manuscript.
Additional support for this work was provided by NSF grants
AST 98-19694 and AST 02-05427 (to M. S.).
This research has made use of the SIMBAD database, operated
at CDS, Strasbourg, France.
The authors wish to extend special thanks to those
of Hawaiian ancestry on whose sacred mountain we are
privileged to be guests.

\appendix

\section{CIT Color Transformations}

In order to present the QUIRC photometry in the standard CIT system,
we derived a set of transformation equations based on data from
\citet{hum84}, \citet{leg93}, and Rayner
(1999, private communication).  We make the assumptions that 
(1) the color response of the QUIRC camera, with a
1024$^2$ HgCdTe HAWAII detector, on the UH 2.2 m
was identical to that of the previous facility camera with a
NICMOS-3 detector and, to the best of our knowledge,
identical filters, and (2) that the NASA IRTF RC1 and RC2
InSb photometers also manifested identical color response
to each other.

$K_{CIT} = 0.959 K_{QUIRC} + 0.041 H_{QUIRC}$

$H_{CIT} = -0.112 K_{QUIRC} + 1.112 H_{QUIRC}$

$J_{CIT} = 0.847 J_{QUIRC} + 0.360 H_{QUIRC} - 0.207 K_{QUIRC}$

\section{Notes on Individual Systems}

\noindent HBC 248

The smallest mass ratio in the sample, q$=$0.04, was observed in this system.  
\citet{hug94} identified HBC 248, unresolved,
as a K2, the same as our identification for the A component.
We found a larger primary star
$A_v$, 4.0 mag, than \citet{hug94} detected for the pair,
1.5 mag; our estimate for the luminosity of the system is about
twice as great as theirs.
The secondary star $A_v$ appears to be close to zero, differing
from the primary $A_v$ at the 2 $\sigma$ level.  In contrast to the results
of \citet{geo01}, our data indicate that
the K2 primary and M6 secondary are coeval
within the 1 $\sigma$ uncertainties.  \citet{geo01} use a fit to
the spectral energy distribution to assign a secondary $T_{eff}$
of $\sim$4170 K, corresponding to a K7.  However,
based on Figures 1 and 2, HBC 248 B
appears to have a late M spectral type.
The small $K$-band excesses in the HBC 248 components are insignificant,
given the uncertainties, consistent with the small $K-L$ color excesses
(Table 9) and indicating an absence of detected circumstellar material.

\bigskip

\noindent HBC 620 

This system, comprised of an M0 primary and an M4.5 secondary,
is coeval and appears to suffer little or no
extinction.  The $K$-band excess is effectively zero in both
components, and the unresolved $K-L$ color is photospheric.  Therefore,
since no color evidence for a circumstellar disk is detected, the
small Br$\gamma$ line flux present in the secondary (Table 9) is
probably attributable to chromospheric activity
in this late M-type star \citep{wal94}.
Within the 1 $\sigma$ uncertainties, our results for the spectral
type and $A_v$ of the primary agree with those of \citet{hug94};
our value for the
total luminosity of the system is about 50 \% greater.

\bigskip

\noindent HBC 625

As in the case of HBC 620, the component stars are coeval, unextinguished,
and present no significant $K$-band excess or $K-L$ color excess.
The primary star spectral type and lack of extinction and excess
are consistent with the results of \citet{hug94}, although our value
for the total luminosity is about twice as great.
\citet{bra97} list the spectral types of the stars in this system
as M1 and M3, similar to our estimates of M1.5 and M4.
\citet{kor02} found evidence that the secondary star in this system is
actually a $\sim$37 milliarcsecond pair.

\bigskip

\noindent AS 205

Falling on the 
1$\times$10$^5$ year isochrone, the coeval components of AS 205 are among the
youngest observed in this sample.  Both stars are veiled, in particular the   
primary, and both spectra display strong Br$\gamma$ emission lines,
indicative of active accretion, and $K-L$ colors
indicative of circumstellar material.  Our estimates of the AS 205 A
$A_v$ and $L$ agree well with those of \citet{coh79}.  For the primary
and secondary spectral types, \citet{coh79} list K0 and K5, while we
find K5 and M3.  Because of the large degree of veiling,
these results have relatively high uncertainties.  From the spectra
of the VV CrA components and AS 353 A, we know that NaI and CO
ca occur in emission.  In 
conjunction with continuum veiling, line emission from hot gas
partially filling in absorption line
spectra could account for the greater difficulty in
fitting the AS 205 spectra to standard stars.  Furthermore,
\citet{kor02} found that the AS 205 secondary is probably a $\sim$9
milliarcsecond double.  This might account in part for the difficultly
in fitting AS 205 B to a single, spectral type standard.
Along with the emission line object, AS 353 A (Figure 2),
the AS 205 primary is located
more than 1 $\sigma$ {\it below} the CTTs locus shown in Figure 3,
possibly the result of the colors being effected by bound-free and
free-free emission or scattering \citep{lad92}.

\bigskip

\noindent WSB 4

Please see \S 4.3.2 for additional details on WSB 4.
We find a spectral type of M3 for both components in this coeval system,
similar to the results of \citet{mar98}, M3 and M3.5 for the
primary and secondary, respectively.
Extinction is negligible along the line of sight to both
stars; the south-east component, WSB 4 B, displays
a small amount of veiling, indicated by a $K$-band excess.  WSB 4 B is
one of the few stars in our sample for which a linear excess with a
small, positive slope yielded the best fit (\S 3.2). 

\noindent The $K-L$ color of WSB 4 A is within $\sim$1 $\sigma$
of photospheric, whereas the B component
has a significant color excess.  \citet{mar98} argues that the 20 \AA~
equivalent width noted by \citet{mey93} in WSB 4 B is
attributable to chromospheric activity, and not to the 
classical T Tauri nature of the system.  It is more probable
that there is still some accretion
occurring and an optically thick, circumstellar disk
present around the secondary.  A growing inner gap in the disk
would account for the lack of IR excess at wavelengths
shorter than the $K$-band (e.g., Haisch et al. 2001) and the 
relatively modest Br$\gamma$ emission line flux.

\bigskip

\noindent WSB 19

The properties we derived for the coeval components
in this weak-lined system are very similar to the results
of \citet{bra97}.  In particular, values for
the $A_v$, $\sim2-3$ mag, are almost
identical and the spectral types are consistent to within one
and a half subclasses.  The component luminosities
of \citet{bra97} are both about three times smaller than what
we derive.  Hence, although both sets of mass estimates agree
well, the ages we derive are younger, $\la$1$\times$10$^6$ years
compared to $\sim$3$\times$10$^6$ years.  

We find a
significant, positively-sloped (i.e. greater at longer wavelengths)
$K$-band excess for the primary, suggestive of circumstellar material.
In the color-color diagram, 
WSB 19 A falls on the line between large and small near-IR excesses.
This system is an IRAS source.  Using the 12 $\mu$m value for
the IRAS flux density, 0.421 Jy, we find a value for the
total $K-12 \mu m$ color of $\sim$4, suggestive of an
optically thick, circumstellar
disk at a radius of $\la$1 AU \citep{skr90, sp95}
around  at least the primary, given its $K$-band excess (Table 7).
No $L$-band data was available for WSB 19.

\bigskip

\noindent WSB 28

The M7 secondary star in WSB 28 is the lowest mass object found in our
sample.  From its location on the H-R diagram (Figures 6 and 7), it
appears to lie below the substellar boundary.  From the tracks of
\citet{pal99}, we obtain only an upper limit of 0.1 M$_{\odot}$,
however, using the \citet{bar98} tracks in Figure 6
we estimate a mass of 0.06 M$_{\odot}$.  There is no evidence
for strong veiling, $K-L$ excesses, or hydrogen line emission in this
system, although the WSB 28 B $K$-band excess was found to have a
linear, slightly positive slope, similar to that of WSB 4 A, but with a
much smaller amplitude (by a factor of 25).
The components appear to be coeval on the \citet{bar98}
tracks and by extrapolation on the \citet{pal99} tracks.  The binary
separation, 5$\farcs$1, is among the widest in the sample.  The
primary star is an M3.

\noindent The values from \citet{bon01} for the
unresolved $A_v$ and luminosity are 4.3 mag and
0.36 L$_{\odot}$, respectively.  We find a total luminosity of about
twice this value.  The primary and secondary star $A_v$ values which
we measure, 5.1$\pm$0.6 and 2.5$\pm$1, respectively, differ
both from each other and from the results of \citet{bon01}.

\bigskip

\noindent DoAr 24E

DoAr 24E B is an infrared companion with no detectable
photospheric features (Figure 2) and a
large $K-L$ color.  In fitting the DoAr 24E A spectrum, a range of spectral 
types, from K0 to K2, were identified.  Therefore, the designated
spectral type, K0, similar
to the K1 found by \citet{luh99}, is assigned
a large uncertainty in Table 7.  The $A_v$, $\sim$6 mag,
agrees well with values reported by \citet{luh99} and
\citet{bon01}.  Our $\sim$9 L$_{\odot}$ 
luminosity for DoAr 24E A agrees well with the 10 L$_{\odot}$ found
by \citet{luh99}.
The modest veiling and $K-L$ color excess (Tables
7 and 9) imply the presence of circumstellar material around the
primary, although its evolutionary stage appears to be
different from that of DoAr 24E B, since the photosphere
of the latter is unobservable.  This discrepancy may be
attributable to the geometry of
this system, i.e., the secondary may be oriented such that it is
observed through its optically thick disk.  Data of \citet{kor02} suggest
that DoAr 24E B is a $\sim$6 milliarcsecond pair.

Although it is not indicated in Figure 2, DoAr 24 E B appears
to have weak HeI 2.058 $\mu$m line emission of approximated the
same equivalent width as its Br$\gamma$ emission line.  We interpret
the HeI emission as a signature of an accretion shock in optically
thick gas.  The excitation conditions which produce this transition
are complex \citep{sim84, geb84}.

\bigskip

\noindent DoAr 26

Little is known about this 2$\farcs$3 separation pair.  We detect modest
extinction towards both components, $A_V \sim$3 and 1 for the primary and
secondary, respectively.
The primary star, an M4, displays strong veiling, consistent with its location
on the color-color diagram.  The secondary, an M6, shows no evidence of
circumstellar material.  The DoAr 26 components are coeval.

\bigskip

\noindent ROX 15

We find later spectral types for the coeval
ROX 15 components, M3 for both, than those
identified by \citet{luh99} and \citet{gre95} for the unresolved
systems, M0 in each case.  Our estimates for the component $A_v$'s are
comparable to the unresolved values in \citet{luh99} and \citet{gre95}.
Our estimate of the total
luminosity, 3.80$\pm$0.99 L$_{\odot}$, is consistent
with that of \citet{luh99}, 3.5 L$_{\odot}$.  Modest veiling is
observed, but no evidence of a $K-L$ color excess or accretion
activity is seen in this system.

\bigskip

\noindent SR 21

The components of SR 21 do not appear to be coeval.  Our estimates
for the spectral type of the primary are somewhat uncertain as hydrogen
emission may fill in the Br$\gamma$ absorption line, the strongest
feature in the spectrum, rendering its use problematic.  Hence, we have
assigned a large uncertainty to the spectral type in Table 7.
The possibility of accretion activity is suggested by the strong
veiling and large $K-L$ color measured for SR 21 A.
Our measurements of the $A_v$ agree with those of \citet{luh99} for
SR 21 A, however they find a luminosity almost twice
as large as ours, probably because of their much earlier
spectral type, F4, estimated for the primary star.  In contrast,
\citet{wil89} estimate a bolometric luminosity of 7 $L_{\odot}$,
only $\sim$25 \% of what we measure.  The secondary star, an M4,
at a separation of 6$\farcs$4, appears to be unremarkable, however,
it is one of the younger objects in the sample at an age of
1$\times$10$^5$ years.  This may be an indication that its luminosity
was overestimated.  There is no evidence for circumstellar 
material around SR 21 B.

\bigskip

\noindent YLW 15A

This is the only target in this study which is not seen at
visible wavelengths. With a separation of 7$\farcs$6, it is the widest   
binary in the sample.  The $K$-band flux ratio of the 2 components   
is about 16; it was not possible to observe the secondary star at sufficiently
high signal to noise to enable reliable detection of any spectral features.   
In the following discussion, we will use the designation YLW 15A to refer
to the primary.

\noindent The spectrum of YLW 15A rises dramatically across the
$K$-band; the appearance of very weak absorption features, and the
IR nature of this source, suggest that
the photosphere was observed under a large visual extinction.
Although we did not detect this source in the $J$-band, \citet{all02}
observed it with the $HST$/NICMOS Camera 3 and found F110W and
F160W magnitudes of 18.22 $\pm$0.16 and 13.05 $\pm$0.01,
respectively.  Using their
equations to transform to CIT $J$- and $H$-band magnitudes, we
find $J=$17.44 $\pm$0.15 mag and $H=$12.76 $\pm$0.02 mag.
Because the S/N ratio of the $H$-band magnitude is higher for the
data of \citet{all02} than in ours, we combined these $J$- and $H$-band
values together with our measurement 
of $K=$9.72 $\pm$0.02, and found an $A_v$ of 32 mag.
We note that our value of $H$, 14.2 mag, is significantly different
from that of \citet{all02}, indicating an appreciable degree of
veiling in this source (e.g., Greene \& Lada 2002).
Applying our spectral approach to finding
the best values of $A_v$, spectral type, and
r$_k$ (measured at 2.2 $\mu$m) yielded 38 mag, K2, and 1.7, respectively.  
Given the veiling, r$_k=$1.7, if we assume the
presence of circumstellar material, implying that the
dereddened $K-L$ color is $>$0.4,
then the upper limit on the extinction is 46 mag,
very close to the upper limit defined for the $A_v$ of YLW 15A by
\citet{wil89} of 44 mag.  \citet{gre02} determined a K5 spectral type
based on high resolution $K$-band spectroscopy.  Our estimate
of the luminosity of YLW 15A, 11.4 L$_{\odot}$,
agrees with that of \citet{wil89} to within 1 $\sigma$.
Figure 14 shows the spectrum of YLW 15A after accounting for 38 magnitudes of
extinction and $r_k=1.7$, in comparison to a K2 spectral type standard.
YLW 15A is one of the younger and more luminous sources in our
sample.

\bigskip

\noindent WSB 71

The components of WSB 71 provided one of the most striking examples
of stars which could be well fit to a range of spectral type standards
using a range of values for $A_v$ and r$_k$.  The uncertainties in the
spectral type determination (Table 7) reflect this ambiguity.
WSB 71 A, a K2, displays
strong veiling.  A large $K-L$ color excess provides evidence
of circumstellar material.  However, only in the secondary, an M6 star, which 
is not veiled and shows little if any color excess, is there an emission
line.  This may be just the result of chromospheric activity.  The total
luminosity of this system, 2.64 L$_{\odot}$, is consistent with that
found by \citet{gre94}, 2.1 L$_{\odot}$.
The stars in this system are not coeval, the primary and secondary
extinctions do
not agree, only one component shows evidence of a disk, and the
separation is large ($\sim$5$''$).  
Proper motion studies would help to confirm this system as a true
binary.  \citet{kor02}, using very high resolution
speckle holography, found evidence for a dusty halo around the 
primary star.  Some of the difficulty we encountered in characterizing
this system may be attributable to related effects from this structure.

\bigskip

\noindent S CrA

Both components of S CrA, a K3 and an M0, are strongly veiled, display large
$K-L$ colors, and have large Br$\gamma$ line fluxes, all 
indicative of active accretion from circumstellar disks.
A linear fit with a positive slope was necessary to define a
satisfactory $K$-band excess.  We determined a total luminosity
of $\sim$3 L$_{\odot}$, a somewhat lower value than that found
by \citet{wil92}, 5.5 L$_{\odot}$.
The parameters for the stars are consistent;
the two components appear to be coeval.  

\bigskip

\noindent HBC 679

HBC 679 A and B, a K5 and an M3, respectively,
appear to be coeval.  Neither component displays
any evidence of circumstellar material.  The $A_v$ of the primary
is about 2 $\sigma$ greater than that of the secondary, however,
since this is a wider binary (separation $\sim$5$''$; 650 AU), this may
be an environmental effect.

\bigskip

\noindent VV CrA

When this system was observed in both 1996 and 1997, the 2 components   
displayed a variety of strong emission lines, including H$_2$ in the   
secondary and the 2.058 $\mu$m He I line in the primary.
We observed this system twice,
primarily in order to verify the He I detection.
In the 1996 observations, the primary spectrum appears to
display a small Ca I line in absorption.  The   
presence of Ca I absorption in the stellar photosphere implies a   
spectral type of about K2 or later, consistent with the characterization   
of the system as a K star by \citet{app86}.  The   
He I emission line requires a powerful ionizing source, probably
in the form of an accretion shock in optically thick gas. 

The continuum flux of the secondary star appears to be varying with   
respect to that of the primary.  For the 1996 observations, the secondary   
to primary flux ratio at 2.2 $\mu$m was 1, whereas it dropped to   
$\sim$0.6 in 1997.  This is consistent with the variable behavior   
observed in this system between 1993 and 1996 \citep{pra97}.

\noindent AS 353
   
As in the case of DoAr 24E, the components of the AS 353 system appear   
to be in distinct evolutionary stages.  In this   
case, the primary star exhibits no absorption lines whatsoever; rather,
the spectrum shows a variety of   
strong emission features, including the 2.058 $\mu$m line of He I
(see above discussion of DoAr 24 E and VV CrA).
Repeated observations of the primary one   
year later (not shown in Fig. 2) produced an almost identical
emission line spectrum.  In contrast, the   
secondary component appears to be a normal, WTT M3 star with
absorption lines, no veiling, and a small
$K-L$ color excess.  Our spectral type for AS 353 B is later 
than that found by \citet{coh79}, M0, but our estimate for $A_v$
is equivalent to theirs within 1$\sigma$.

\noindent The distance to this system is uncertain.  \citet{dam85}
cite 200$\pm$100 pc to the Aquila Rift,
which provides an upper limit for the distance to the
stars if we assume that the AS 353 binary
is in front of the dark cloud complex.  
\citet{edw82} suggest a distance of 150 pc, assuming AS 353 is
in Gould's Belt.  We adopt d$=$150$\pm$50 pc.
This is consistent with a location
of AS 353 B on the H-R diagram which implies an age of 5$\times10^5$ years
for this star.  Given the unremarkable appearance of this star, its
young age seems incongruous.  \citet{whi03} find that AS 353 B is itself a
close binary pair with an 0$\farcs$2 separation.  If we assume a
luminosity ratio of 1.0, then each component of AS 353 B has a
luminosity of 0.34 L$_{\odot}$, implying a more reasonable age of 
1$\times$10$^6$ years.

\noindent Given the disparate appearance 
of the AS 353 components, i.e. absorption {\it versus}
emission line spectra, and the relatively 
large separation, 5$\farcs$7, of
this system, it is possible that AS 353 is not a true binary.
$\sim$25 background objects were observed in
our 1'$\times$1' QUIRC field, yielding a field star density of
$\sim7\times10^{-3}$ stars per square arcsecond, 
about 140 times greater than that measured by
\citet{gre92} for Ophiuchus.  Common proper motion measurements
of the components of AS 353 would help to confirm the stars'
relationship.

\clearpage

\pagestyle{empty}

\begin{deluxetable}{llllllrcl}
\rotate
\tablewidth{0pt}
\tablecaption{Spectroscopic Sample \label{tbl-1}}
\tablehead{
\colhead{$  $} & \colhead{$  $} & \colhead{R.A.} & \colhead{Decl.} & \colhead{$K$} & \colhead{Sep.} & \colhead{P.A.} & \colhead{Year} & \colhead{$  $}\\
\colhead{Source Name(s)} & \colhead{$  $} & \colhead{(J2000.0)} & \colhead{(J2000.0)} & \colhead{(mag)} & \colhead{($''$)} & \colhead{($^{\circ}$)} & \colhead{Observed} & \colhead{SFR}}
\startdata
HBC 248      & Sz 68      & 15 45 12.9 & $-$34 17 30.6 & 6.52\tablenotemark{a} & 2.6 & 295 &
1997     & Lup \\
HBC 620      & Sz 108     & 16 08 42.7 & $-$39 06 18.3 & 8.63\tablenotemark{a} & 4.2 &  25 &
1997     & Lup \\
HBC 625      & Sz 116     & 16 09 42.6 & $-$39 19 42.0 & 9.48\tablenotemark{a} & 1.5 &  29 &
1997     & Lup \\
AS 205       & HBC 254    & 16 11 31.40 & $-$18 38 24.5 & 5.54\tablenotemark{b} & 1.3 & 204 &
1996     & Oph \\
WSB 4        &   $...$   & 16 18 50.3 & $-$26 10 08.0 & 9.59       & 2.8 & 129 &
1997     & Oph \\
WSB 19       &   $...$    & 16 25 02.2 & $-$24 59 31.0 & 9.12       & 1.5 & 264 &
1997     & Oph \\
WSB 28       & ISO$-$Oph 27 & 16 26 20.7 & $-$24 08 48.0 & 9.38       & 5.1 & 358 &
1997     & Oph \\
DoAr 24E     &  HBC 639   & 16 26 23.4 & $-$24 21 02.0 & 6.7    & 2.2 & 150 &
1996     & Oph \\
DoAr 26      &  WSB 35    & 16 26 34.8 & $-$23 45 41.0 & 8.89       & 2.3 & 132 &
1996     & Oph \\
ROX 15       &Elias 2-26  & 16 26 42.9 & $-$24 20 32.0 & 8.00       & 1.2 &  70 &
1996     & Oph \\
SR 21        & YLW 8   & 16 27 10.2 & $-$24 19 16.0 & 6.35   & 6.4 & 175 &
1996     & Oph \\
YLW 15A      & ISO$-$Oph 141 & 16 27 27.1 & $-$24 40 51.0 & 9.65   & 7.6 &  21 &
1997     & Oph \\
WSB 71 & H$\alpha$ 71     & 16 31 30.9 & $-$24 24 40.0 & 8.1   & 4.8 &  40 &
1996     & Oph \\
S CrA        & HBC 286    & 19 01 08.7 & $-$36 57 19.8 & 6.08       & 1.3 & 160 &
1996     & CrA \\
HBC 679      & WaCrA3     & 19 02 22.4 & $-$36 55 41.0 & 9.09       & 4.5 &  60 &
1996     & CrA \\
VV CrA       & HBC 291    & 19 03 06.7 & $-$37 12 51.0 & 6.49       & 1.9 &  45 &
1996, 1997 & CrA \\
AS 353       & HBC 292    & 19 20 31.0 & $+$11 01 54.9 & 7.73       & 5.7 & 175 &
1996     & L673\\
\enddata

\tablecomments{Units of right ascension are hours, minutes, and seconds, and units of declination are degrees, arcminutes, and arcseconds.}

\tablenotetext{a}{Hughes et al. 1994}
\tablenotetext{b}{Liu et al. 1996}

\end{deluxetable}

\clearpage

\pagestyle{empty}

\begin{deluxetable}{lll}
\tablewidth{0pt}
\tablecaption{Spectral Type Standards \label{tbl-2}}
\tablehead{
\colhead{$  $} & \colhead{Spectral} & \colhead{$K$}\\
\colhead{Name} & \colhead{Type} & \colhead{(mag)}}
\startdata
HR 88    &  G2.5 V   &  4.9 \\
HR 8631  &  G4 V     &  4.2 \\
HR 9088  &  G5 Vb    &  4.2 \\
HR 8784  &  G8 IV    &  4.6 \\
HR 166   &  K0 V     &  4.0 \\
GL 28    &  K2 V     &  5.1 \\
HR 8832  &  K3 V     &  3.1 \\
HR 8085  &  K5 V     &  2.4 \\
HR 8086  &  K7 V     &  2.9 \\
GL 763   &  M0 V     &  5.9 \\
GL 752A  &  M3 V     &  4.7 \\
GL 791.2 &  M4.5 V   &  7.3 \\
GL 83.1  &  M5 V     &  6.7 \\
GL 644C  &  M7 V     &  8.8 \\
GL 752B  &  M8 V     &  8.8 \\
\enddata

\end{deluxetable}

\clearpage

\pagestyle{empty}

\begin{deluxetable}{llccccl}
\tablewidth{0pt}
\tablecaption{Component Resolved Photometry \label{tbl-3}}
\tablehead{
\colhead{$  $} & \colhead{$  $} & \colhead{$J$} & \colhead{$H$} & \colhead{$K$} &
\colhead{$L$} & \colhead{References}\\
\colhead{Name} & \colhead{$  $} & \colhead{(mag)} & \colhead{(mag)} & \colhead{(mag)} &
\colhead{(mag)} & \colhead{($L$-band)}}
\startdata
HBC 248\tablenotemark{a}& A  & $ 7.67 \pm 0.03 $ & $ 6.96 \pm 0.08 $ & $ 6.62 \pm 0.09 $ & $ 6.07 \pm 0.05 $ & 1 \\
  & B         & $10.58 \pm 0.03 $ & $ 9.78 \pm 0.10 $ & $ 9.29 \pm 0.09 $& $8.81 \pm 0.08 $ & 1 \\
HBC 620\tablenotemark{a}& A  & $ 9.76 \pm 0.02 $ & $ 9.09 \pm 0.02 $ & $ 8.78 \pm 0.01 $ & $L_{total}=8.48 \pm 0.01$ & 2 \\
  & B      & $12.02 \pm 0.04 $ & $11.21 \pm 0.03 $ & $10.86 \pm 0.03 $& $  $ & $  $ \\
HBC 625\tablenotemark{a}& A  & $10.84 \pm 0.03 $ & $10.13 \pm 0.01 $ & $ 9.91 \pm 0.02 $ & $L_{total}=9.36 \pm 0.02$ & 2 \\
  & B       & $11.48 \pm 0.03 $ & $10.90 \pm 0.01 $ & $10.71 \pm 0.02 $& $  $ & $  $ \\
AS 205\tablenotemark{b}& A  & $ 8.15 \pm 0.06 $ & $ 7.16 \pm 0.06 $ & $ 5.90 \pm 0.03 $ & $4.9 \pm 0.1 $ & 3, 4 \\
  & B      & $ 9.24 \pm 0.17 $ & $ 7.97 \pm 0.13 $ & $ 6.90 \pm 0.14 $& $ 5.8 \pm 0.1 $ & 3, 4 \\
WSB 4 & A         & $11.70 \pm 0.06 $ & $10.96 \pm 0.04 $ & $10.63 \pm 0.06 $& $ 9.50 \pm 0.12 $ & 5 \\
  & B             & $11.55 \pm 0.06 $ & $10.69 \pm 0.03 $ & $10.12 \pm 0.06 $& $ 10.21 \pm 0.12 $ & 5 \\
WSB 18\tablenotemark{c}& A  & $11.47 \pm 0.05 $ & $10.48 \pm 0.03 $ & $10.01 \pm 0.06 $ & $...$ & $  $ \\
  & B      & $12.18 \pm 0.06 $ & $11.11 \pm 0.03 $ & $10.34 \pm 0.06 $& $...$ & $  $ \\
WSB 19 &  A      & $11.16 \pm 0.03 $ & $10.13 \pm 0.04 $ & $ 9.53 \pm 0.15 $& $...$ & $  $ \\
  & B            & $11.76 \pm 0.04 $ & $10.85 \pm 0.04 $ & $10.38 \pm 0.16 $& $...$ & $  $ \\
HBC 257\tablenotemark{c}& A   & $ 9.45 \pm 0.07 $ & $ 8.46 \pm 0.03 $ & $ 7.86 \pm 0.09 $ & $...$ & $  $ \\
  & B       & $10.17 \pm 0.08 $ & $ 9.29 \pm 0.05 $ & $ 8.89 \pm 0.12 $& $...$ & $  $ \\
WSB 26\tablenotemark{c}& A  & $11.45 \pm 0.06 $ & $10.45 \pm 0.04 $ & $ 9.46 \pm 0.08 $ & $...$ & $  $ \\
  & B       & $11.63 \pm 0.06 $ & $10.46 \pm 0.04 $ & $ 9.78 \pm 0.08 $& $...$ & $  $ \\
WSB 28 & A              & $10.89 \pm 0.06 $ & $ 9.82 \pm 0.03 $ & $ 9.48 \pm 0.07 $& $...$ & $  $ \\
  & B     & $13.41 \pm 0.18 $ & $12.37 \pm 0.11 $ & $11.84 \pm 0.18 $& $...$ & $  $ \\
DoAr 24E\tablenotemark{d}& A  & $ 9.2  \pm 0.3  $ & $ 7.8  \pm 0.2  $ & $ 7.1  \pm 0.1  $ & $ 6.43 \pm 0.07 $ &6 \\
  & B     & $12.1  \pm 0.4  $ & $ 9.7  \pm 0.2  $ & $ 8.1  \pm 0.1  $& $ 5.9 \pm 0.1   $ & 6 \\
DoAr 26 & A             & $11.17 \pm 0.07 $ & $ 9.98 \pm 0.04 $ & $ 9.17 \pm 0.06 $& $...$ & $  $ \\
  & B       & $11.83 \pm 0.09 $ & $10.98 \pm 0.07 $ & $10.56 \pm 0.08 $& $...$ & $  $ \\
ROX 15 & A              & $10.70 \pm 0.04 $ & $ 9.19 \pm 0.05 $ & $ 8.29 \pm 0.05 $& $L_{total}=7.09 \pm 0.10$ &7 \\
  & B       & $12.20 \pm 0.07 $ & $10.57 \pm 0.09 $ & $ 9.66 \pm 0.07 $& $               $ & $  $\\
SR 21\tablenotemark{d}& A  & $ 8.77 \pm 0.02 $ & $ 7.45 \pm 0.04 $ & $ 6.40 \pm 0.03 $ & $ 5.25 \pm 0.03 $ & 6\\
  & B       & $12.03 \pm 0.04 $ & $10.54 \pm 0.03 $ & $ 9.70 \pm 0.02 $& $ 9.26 \pm 0.04 $ & 6\\
YLW 15A\tablenotemark{d}& A  &  $...$  & $ 14.2 \pm 0.1  $ & $ 9.72 \pm 0.02 $ & $ 6.70 \pm 0.01 $ & 6\\
  & B      &  $...$ & $ 15.7 \pm 0.2  $ & $12.69 \pm 0.06 $ & $ 11.0 \pm 0.5  $ & 6\\
WSB 71\tablenotemark{d}& A & $11.00 \pm 0.03 $ & $ 9.50 \pm 0.03 $ & $  8.3 \pm 0.2  $ &$ 6.67 \pm 0.01$&6 \\
  & B      & $11.4  \pm 0.2  $ & $10.34 \pm 0.2  $ & $ 10.0 \pm 0.2  $& $ 9.17 \pm 0.04 $ & 6\\
Elias 2-49\tablenotemark{c}& A & $ 6.99 \pm 0.03 $ & $ 6.18 \pm 0.05 $ & $ 5.32 \pm 0.04 $ & $...$ & $  $ \\
  & B      & $ 9.00 \pm 0.09 $ & $ 8.14 \pm 0.12 $ & $ 7.73 \pm 0.10 $& $...$ & $  $ \\
S CrA & A               & $ 8.60 \pm 0.04 $ & $ 7.50 \pm 0.05 $ & $ 6.56 \pm 0.06 $& $ 5.06\tablenotemark{e} $ & 2, 9 \\
  & B            & $ 9.37 \pm 0.06 $ & $ 8.28 \pm 0.08 $ & $ 7.27 \pm 0.08 $& $ 6.07\tablenotemark{e} $ & 2, 9 \\
HBC 679 & A             & $10.37 \pm 0.03 $ & $ 9.50 \pm 0.04 $ & $ 9.24 \pm 0.06 $& $ L_{total}=9.12 \pm 0.04 $&10 \\
  & B         & $12.48 \pm 0.07 $ & $11.56 \pm 0.06 $ & $11.18 \pm 0.08 $& $  $ & $  $ \\
VV CrA & A              & $ 9.60 \pm 0.05 $ & $ 8.38 \pm 0.05 $ & $ 7.27 \pm 0.05 $& $ 6.28 \pm 0.04 $ & 2, 11 \\
  & B         & $12.07 \pm 0.22 $ & $ 9.69 \pm 0.08 $ & $ 7.33 \pm 0.05 $& $ 3.70 \pm 0.04 $ & 2, 11\\
AS 353 & A              & $ 10.03 \pm 0.03 $ & $ 9.16 \pm 0.02 $ & $ 8.23 \pm 0.03 $& $ 6.35 \pm 0.01 $ & 8 \\
  & B          & $10.22 \pm 0.04 $ & $ 9.20 \pm 0.04 $ & $ 8.76 \pm 0.04 $& $ 8.21 \pm 0.08 $ & 8 \\
\enddata

\tablecomments{All $J$-, $H$-, and $K$-band QUIRC and IRTF photometry is presented in the CIT system.}

\tablenotetext{a}{$J$-, $H$-, and $K$-band photometry from Hughes et al. (1994)}
\tablenotetext{b}{$J$-, $H$-, and $K$-band photometry from Liu et al. (1996)}
\tablenotetext{c}{Not members of the spectroscopic sample.}
\tablenotetext{d}{Photometry from IRTF.}
\tablenotetext{e}{No uncertainty was given in the references.}

\tablerefs{
(1) Geoffray \& Monin 2001;
(2) Chelli et al. 1995;
(3) Hughes et al. 1994;
(4) Cohen 1974;
(5) Meyer et al. 1993;
(6) this work;
(7) Greene et al. 1994;
(8) Cohen \& Schwartz 1983;
(9) Knacke et al. 1973;
(10) Walter et al. 1997;
(11) Wilking et al. 1992.}

\end{deluxetable}

\clearpage

\pagestyle{empty}

\begin{deluxetable}{lcccc}
\tablewidth{0pt}
\tablecaption{Spectral Features and Bandpasses for Equivalent Widths \label{tbl-4}}
\tablehead{
\colhead{$\lambda_{central}$} & \colhead{$\Delta\lambda$} & \colhead{$  $} & \colhead{$  $} & \colhead{$  $}\\
\colhead{($\mu$m)} & \colhead{(\AA)} & \colhead{Species} & \colhead{Transition} & \colhead{Comments}}
\startdata
2.05869 & 120 & He I    & $2p^{1}P^{o} - 2s^{1}S$                     & $\oplus$ Contamination \\
2.10655 & 85 & Mg I    & $4f^{1}F^{o}_{3} - 7g^{1}G_{4}$             & blend                  \\
2.10666 & 85 & Mg I    & $4f^{3}F^{o}_{2,3,4} - 7g^{3}G^{o}_{3,4,5}$ & blend                  \\
2.10988 & 85 & Al I    & $4p^{2}P^{o}_{1/2} - 5s^{2}S_{1/2}$         & blend          \\
2.11695 & 75 & Al I    & $4p^{2}P^{o}_{3/2} - 5s^{2}S_{1/2}$         &
                   \\
2.12183 & 120 & H$_{2}$ & $\nu = 1 - 0\ S(1)$                         &
                   \\
2.16611 & 120 & H I     & $ n = 7 - 4\quad ({\rm Br}\gamma )$         &
                   \\
2.20624 & 70 & Na I    & $4p^{2}P^{o}_{3/2} - 4s^{2}S_{1/2}$         & blend                  \\
2.20897 & 70 & Na I    & $4p^{2}P^{o}_{1/2} - 4s^{2}S_{1/2}$         & blend                  \\
2.26141 & 110 & Ca I    & $4f^{3}F^{o}_{2} - 4d^{3}D_{1}$             & blend \\
2.26311 & 110 & Ca I    & $4f^{3}F^{o}_{3} - 4d^{3}D_{2}$             & blend                  \\
2.26573 & 110 & Ca I    & $4f^{3}F^{o}_{4} - 4d^{3}D_{3}$             & blend                  \\
2.28141 & 100 & Mg I    & $4d^{3}D_{3,2,1} - 6f^{3}F^{o}_{2,3,4}$     &
                   \\
2.29353 & 130 & CO      & $v = 2 - 0\ \rm bandhead$                   &
                   \\
2.32265 & 130 & CO      & $v = 3 - 1\ \rm bandhead$                   &
                   \\
2.35246 & 130 & CO      & $v = 4 - 2\ \rm bandhead$                   &
                   \\
2.38295 & 130 & CO      & $v = 5 - 3\ \rm bandhead$                   &
                   \\
\enddata

\tablecomments{For the Mg I $+$ Al I blend at 2.11$\mu$m, $\Delta\lambda$
refers to the entire blended feature.  The same applies for the bandpasses
of the Na I doublet (2.21$\mu$m) and the Ca I triplet (2.26$\mu$m) blends.}

\end{deluxetable}

\clearpage

\pagestyle{empty}

\begin{deluxetable}{lccccccccccc}
\rotate
\tablewidth{0pt}
\tablecaption{Standard Star $K$ band Equivalent Widths (\AA)\label{tbl-5}}
\tablehead{
\colhead{$  $} & \colhead{Spectral} & \colhead{2.110}  & \colhead{2.117}  & \colhead{2.166}  & \colhead{2.208}  & \colhead{2.264}  & \colhead{2.281}  & \colhead{2.294} & \colhead{2.323} & \colhead{2.352} & \colhead{2.383} \\
\colhead{Name} & \colhead{Type}     & \colhead{Al~I $+$ Mg I} &  \colhead{Al~I}  & \colhead{H~I} & \colhead{Na~I} & \colhead{Ca~I} & \colhead{Mg~I} & \colhead{CO(2$-$0)} & \colhead{CO(3$-$1)} & \colhead{CO(4$-$2)} & \colhead{CO(5$-$3)}}
\startdata
HR   88  & G2.5 V  & 1.7  & 1.5  & 3.7  & 1.1 &  1.1 &  1.2 &   $...$  &   $...$  &   $...$  &   $...$  \\
HR 8631  & G4 V    & 1.1  & 1.1  & 2.4  & 1.1 &  1.2 &  1.0 &  2.2 &  4.0 &  2.9 &  3.8 \\
HR 9088  & G5 V    & 0.7  & 0.5  & 2.9  & 0.9 &  1.0 &  0.9 &  1.0 &  1.0 &  1.1 &  1.5 \\
HR 8784  & G8 IV   & 1.5  & 1.6  & 2.5  & 1.3 &  1.2 &  1.2 &  4.0 &  4.2 &  5.1 &  5.2 \\
HR  166  & K0 V    & 1.3  & 0.6  & 2.5  & 1.8 &  1.4 &  1.8 &  3.7 &  3.5 &  4.9 &  4.3 \\
GL  28   & K2 V    & 2.4  & 2.1  & 1.3  & 2.1 &  2.7 &  2.3 &  4.4 &  5.6 &  4.9 &  5.7 \\
HR 8832  & K3 V    & 2.4  & 1.9  & 1.1  & 2.2 &  2.9 &  1.6 &  4.0 &  4.1 &  5.0 &  6.0 \\
HR 8085  & K5 V    & 1.4  & 1.4  &  $...$   & 2.2 &  2.9 &  1.1 &  4.1 &  3.8 &  4.9 &  5.2 \\
HR 8086  & K7 V    & 1.8  & 1.8  &  $...$   & 3.1 &  3.5 &  1.1 &  4.8 &  5.2 &  5.3 &  5.3 \\
GL 763   & M0 V    & 1.1  & 1.6  &  $...$   & 2.9 &  3.7 &  1.3 &  4.4 &  4.2 &  4.8 &  4.2 \\
GL 752.a & M3 V    & 0.9  & 1.4  &  $...$   & 4.2 &  4.5 &  0.7 &  4.6 &  4.1 &  4.6 &  4.0 \\
GL 791.2 & M4.5 V  &  $...$   &  $...$   &  $...$   & 4.5 &  2.6 &  0.5 &  6.3 &  4.5 &  5.7 &  4.3 \\
GL  83.1 & M5 V    &  $...$   &  $...$   &  $...$   & 4.1 &  1.8 &  0.5 &  5.1 &  4.2 &  4.8 &  3.7 \\
GL 644c  & M7 V    &  $...$   &  $...$   &  $...$   & 3.7 &  0.8 &  0.8 &  6.8 &  6.1 &  5.3 &  4.5 \\
GL 752.b & M8 V    &  $...$   &  $...$   &  $...$   & 4.4 &  0.7 &  1.0 &  8.6 &  6.6 &  6.4 &  4.8 \\
\enddata

\end{deluxetable}

\clearpage

\pagestyle{empty}

\begin{deluxetable}{llcccccccccccc}
\rotate
\tablecolumns{14}
\tablewidth{0pt}
\tablecaption{Equivalent Widths (\AA) of Spectroscopic Sample \label{tbl-6}}
\tablehead{
\colhead{$  $} & \colhead{$  $} & \colhead{2.059} & \colhead{2.110} & \colhead{2.117} &
\colhead{2.122} & \colhead{2.166} & \colhead{2.208} & \colhead{2.264} &
\colhead{2.281} & \colhead{2.294} & \colhead{2.323} & \colhead{2.352} &
\colhead{2.383}\\
\colhead{Name} & \colhead{$  $} & \colhead{He I} & \colhead{Mg I + Al I} & 
\colhead{Al I} & \colhead{H$_{2}$}&
\colhead{H I} & \colhead{Na I} & \colhead{Ca I} & \colhead{Mg I} & 
\colhead{CO(2-0)} & \colhead{CO(3-1)} & \colhead{CO(4-2)} & \colhead{CO(5-3)}}
\startdata
HBC 248 & A     &$...$&  1.4 \quad   &  1.4 &$...$& $...$&   1.4   &  2.2 &  1.1 &  4.1   &   3.4  &  4.4   &  4.1   \\  
        & B     &$...$&  0.7 \quad   &  0.7 &$...$& $...$&   2.0   &  2.0 &  0.4 &  4.3   &   3.6  &  4.2   &  4.6   \\  
HBC 620 & A     &$...$&  2.1 \quad   &  1.8 &$...$& $...$&   3.7   &  4.6 &  1.0 &  6.5   &   4.3  &  4.4   &  5.2   \\  
        & B     &$...$&$...$ &$...$&$...$&  $-$1.54 &   1.6   &  1.4 &$...$&  3.4   &   1.1  &  3.5   &  2.4   \\  
HBC 625 & A     &$...$&  1.3 \quad   &  1.5 &$...$& $...$&   4.1   &  4.1 &  0.8 &  5.3   &   4.2  &  5.9   &  5.3   \\  
        & B     &$...$&  0.4 \quad   &  0.7 &$...$& $...$&   3.5   &  4.0 &  0.6 &  5.8   &   2.3  &  5.1   &  3.6   \\  
AS 205  & A     &$...$&$...$ &$...$&$...$&  $-$3.0 & $...$&  0.8 &$...$&$...$& $...$ &$...$&$...$\\  
        & B     &$...$&  1.1 \quad   &  0.8 &$...$&  $-$1.7 &   1.6   &  1.8 &  0.8 &  3.1   &   2.3  &  2.7   &  2.6   \\  
WSB 4   & A     &$...$&  1.4 \quad   &  1.7 &$...$& $...$&   3.6   &  4.1 &$...$&  4.9   &   2.5  &  3.6   &  2.5   \\
        & B     &$...$&  1.2 \quad   &  1.0 &$...$&  $-$1.0 &   1.7   &  2.6 &  0.5 &  2.7   &   2.8  &  2.3   &  1.5   \\
WSB 19  & A     &$...$&  1.9 \quad   &  1.3 &$...$&  $-$1.4 &   2.6   &  3.0 &  0.6 &  5.2   &   2.2  &  3.7   &  3.7   \\  
        & B     &$...$&  1.6 \quad   &  1.8 &$...$&  $-$0.7 &   2.1   &  2.6 &  1.0 &  6.2   & $...$ &  2.7   &  3.3   \\  
WSB 28  & A     &$...$&  0.6 \quad   &  0.9 &$...$& $...$&   3.8   &  4.7 &  1.6 &  7.8   &   6.3  &  6.7   &  6.2   \\  
        & B     &$...$&$...$ &$...$&$...$& $...$&   0.8   &  0.7 &  0.6 &  4.7   &   4.8  &  1.9   &  3.4   \\  
DoAr 24E & A     &$...$&  1.3 \quad   &  0.6 &$...$& $...$&   1.2   &  1.4 &  0.3 &  3.9   &   2.6  &  3.6   &  3.3   \\  
        & B     &$...$&$...$ &$...$&$...$&  $-$0.9 & $...$&$...$&$...$&$...$& $...$ &$...$&$...$\\  
DoAr 26 & A     &$...$&$...$ &$...$&$...$&  $-$2.2 &   2.4   &  2.1 &  0.5 &  3.6   &   1.5  &  2.7   &  2.8   \\  
        & B     &$...$&$...$ &$...$&$...$& $...$&   2.4   &  3.1 &  1.0 &  5.1   &   3.1  &  2.8   &  4.3   \\  
ROX 15  & A     &$...$&  1.5 \quad   &  1.4 &$...$& $...$&   2.4   &  2.8 &  0.8 &  5.7   &   4.4  &  4.6   &  5.1   \\
    & B     &$...$&  1.1 \quad   &  1.7 &$...$& $...$&   2.3   &  2.8 &  0.9 &  6.3   &   4.1  &  2.2   &  4.8   \\  
SR 21   & A     &$...$&  0.7 \quad   &  0.5 &$...$&   2.1   &   0.5   &$...$&  0.5 &$...$& $...$ &$...$&$...$\\  
        & B     &$...$&  1.2 \quad   &  0.8 &$...$& $...$&   3.4   &  3.6 &$...$&  6.3   &   5.5  &  4.7   &  5.5   \\  
YLW 15A & A     &$...$&$...$ &$...$&$...$&  $-$1.3 &   0.7   &  0.9 &$...$&$...$& $...$ &$...$&$...$\\  
        & B     &$...$&$...$ &$...$&$...$& $...$& $...$&$...$&$...$&$...$& $...$ &$...$&$...$\\  
WSB 71 & A &$...$&  1.0 \quad   &  0.8 &$...$& $...$&   1.4   &  1.3 &  0.9 &  2.4   &   2.6  &  1.8   &  2.5   \\  
             & B &$...$&$...$ &$...$&$...$&  $-$1.8 &   2.9   &  3.2 &$...$&  6.5   &   4.6  &  4.2   &  5.5   \\  
S CrA   & A     &$...$&$...$ &$...$&$...$&  $-$3.9 & $...$&  0.9 &$...$&  0.9   &   0.5  &  1.1   &  0.9   \\  
        & B     &$...$&$...$ &$...$&$...$&  $-$2.8 & $...$&  0.8 &$...$&  1.3   &   0.5  &  1.1   &  1.0   \\  
HBC 679 & A     &$...$&  2.1 \quad   &  1.7 &$...$& $...$&   3.5   &  4.5 &  1.8 &  4.5   &   4.3  &  5.0   &  4.5   \\  
        & B     &$...$&  1.9 \quad   &  1.4 &$...$& $...$&   3.6   &  3.7 &$...$&  3.5   &   3.5  &  2.3   &  2.6   \\  
VV CrA '96 & A  & $-$2.1      &$...$ &$...$&$...$& $-$15.7 &  $-$0.6 &  2.0 &$...$& $-$5.0 & $-$4.1 & $-$2.3 & $-$1.4 \\  
           & B  &$...$&$...$ &$...$& $-$0.5 & $-$3.5  &  $-$0.5 &  0.7 &$...$& $-$2.3 & $-$2.2 & $-$1.3 & $-$1.3 \\  
VV CrA '97 & A  & $-$1.3      &  0.5 \quad   &  0.2 & $-$0.8 & $-$14.2 &  $-$0.8 &  1.7 &$...$& $-$4.8 & $-$2.9 & $-$2.6 & $-$1.3 \\  
           & B  &$...$&$...$&$...$& $-$2.1 & $-$4.0  &  $-$0.4 &  0.7 &$...$& $-$2.1 & $-$1.5 & $-$2.2 & $-$1.1 \\  
AS 353     & A  &$...$&  0.6 \quad   &  0.6 &$...$& $-$21.1 &  $-$1.0 &  1.2 &$...$& $-$2.4 & $-$0.2 & $-$0.8 &$...$\\  
           & B  &$...$&  1.5 \quad   &  1.6 &$...$& $...$&   2.0   &  3.1 &  0.9 &  6.2   &   6.1  &  4.9   &  5.8   \\ 
\enddata

\end{deluxetable}

\clearpage

\pagestyle{empty}

\begin{deluxetable}{llllllllc}
\tablecolumns{9}
\tablewidth{0pt}
\tablecaption{Derived Sample Properties \label{tbl-7}}
\tablehead{
\colhead{$  $} & \colhead{$  $} & \colhead{Spectral} & \colhead{$T_{eff}$\tablenotemark{a}} & \colhead{$A_V$} & \colhead{$r_k$} & \colhead{$d$\tablenotemark{b}} & \colhead{$L_*$} & \colhead{Reference} \\
\colhead{Name} & \colhead{$  $} & \colhead{Type\tablenotemark{c}} & \colhead{(K)} & \colhead{(mag)} & \colhead{(F$_{K_{ex}}$/F$_{K_*}$)} & \colhead{(pc)} & \colhead{(L$_{\odot}$)} & \colhead{(d)}}
\startdata
HBC 248 & A   & K2 (1)  & 4900$\pm$150 & 4.0$\pm$0.9& 0.10$\pm$0.11 & 159$^{+77}_{-39}$ & 14.49$^{+17.44}_{-6.23}$ &1 \\
        & B   & M6 (1)  & 3000$\pm$150 & 0.0$\pm$1.1& 0.07$\pm$0.18 & 159$^{+77}_{-39}$ & 0.29$^{+0.35}_{-0.12}$ &1 \\
HBC 620 & A   & M0 (1)  & 3800$\pm$200 & 0.4$\pm$1.1  & 0.00$\pm$0.12 & 150 & 0.68$^{+0.19}_{-0.17}$ & 2 \\
        & B   & M4.5(1) & 3200$\pm$150 & 0.5$\pm$1.4 & 0.03$\pm$0.17 & 150 & 0.08$\pm$0.02 & 2 \\
HBC 625 & A   & M1.5(1) & 3650$\pm$150 & 0.0$\pm$0.8 & 0.00$\pm$0.08 & 150 & 0.22$\pm$0.06 & 2 \\
        & B   &  M4 (1) & 3250$\pm$150 & 0.9$\pm$0.7 & 0.03$\pm$0.08 & 150 & 0.11$\pm$0.03 & 2 \\
AS 205  & A   &  K5 (2) & 4450$\pm$400 & 2.9$\pm$1.3  & 2.35$\pm$0.87  & 160 & 7.10$^{+1.89}_{-1.66}$ &3 \\
        & B   &  M3 (2) & 3450$\pm$300 & 2.1$\pm$1.0  & 0.72$\pm$0.15 & 160 & 2.19$^{+0.58}_{-0.51}$ & 3 \\
WSB 4   & A   &  M3 (2) & 3450$\pm$300 & 0.0$\pm$1.0  & 0.00$\pm$0.11 & 160 & 0.11$\pm$0.03 & 3 \\
        & B   &  M3 (1) & 3450$\pm$150 & 0.4$\pm$1  & 0.19$\pm$0.10  & 160 & 0.15$\pm$0.04 & 3 \\
WSB 19  & A   &  M3 (1) & 3450$\pm$150 & 1.7$\pm$1  & 0.37$\pm$0.07  & 160 & 0.29$\pm$0.08 & 3 \\
        & B   &  M5 (1) & 3150$\pm$150 & 2.7$\pm$1.4  & 0.00$\pm$0.15 & 160 & 0.18$\pm$0.05 & 3 \\
WSB 28  & A   &  M3 (1) & 3450$\pm$150 & 5.1$\pm$0.6  & 0.00$\pm$0.06 & 160 & 0.69$^{+0.18}_{-0.16}$ & 3 \\
        & B   &  M7 (1) & 2850$\pm$150 & 2.5$\pm$1  & 0.14$\pm$0.10 & 160 & 0.04$\pm$0.01 & 3 \\
DoAr 24E& A   &  K0 (3) & 5250$\pm$400 & 5.7$\pm$1.0 & 0.25$\pm$0.12 & 160 & 8.78$^{+2.33}_{-2.05}$ & 3 \\
        & B   &  $...$     &   $...$   & $...$   & $...$   &  $...$   & $...$ \\
DoAr 26 & A   &  M4 (1) & 3250$\pm$150 & 3.3$\pm$1.2  & 0.98$\pm$0.49 & 160 & 0.36$^{+0.10}_{-0.08}$ &3 \\
        & B   &  M6 (1) & 3000$\pm$150 & 1.0$\pm$0.9  & 0.14$\pm$0.10 & 160 & 0.11$\pm$0.03 & 3 \\
ROX 15  & A   &  M3 (2) & 3450$\pm$300 & 11.1$\pm$0.8  & 0.06$\pm$0.09 & 160 & 3.45$^{+0.92}_{-0.81}$ & 3 \\
        & B   &  M3 (2) & 3450$\pm$300 & 13.8$\pm$1.1  & 0.22$\pm$0.13 & 160 & 1.35$^{+0.36}_{-0.32}$ & 3\\
SR 21   & A   & G2.5 (3)& 5950$\pm$300 & 9.0$\pm$0.9  & 0.78$\pm$0.54 & 160 & 28.04$^{+7.45}_{-6.57}$ & 3\\
        & B   & M4 (1) & 3250$\pm$150  & 9.3$\pm$1.1  & 0.15$\pm$0.11 & 160 & 0.66$^{+0.18}_{-0.16}$ & 3\\
YLW 15A & A      & K2 (3) & 4900$\pm$500  & 38$\pm$4&1.7$\pm$0.5 & 160 & 11.40$^{+3.03}_{-2.67}$ & 3\\
        & B   & $...$      &   $...$   & $...$   & $...$   &  $...$   & $...$\\
WSB 71 & A & K2 (3) & 4900$\pm$500  & 8.0$\pm$0.7  & 1.12$\pm$0.30 & 160 & 2.30$^{+0.61}_{-0.54}$  & 3\\
        & B   & M6 (2) & 3000$\pm$300  & 4.7$\pm$1.0 & 0.05$\pm$0.12 & 160 & 0.34$\pm$0.09 & 3  \\
S CrA   & A   & K3 (2) & 4800$\pm$400 & 1.0$\pm$1.0 & 2.89$\pm$1.60 & 130 & 2.29$^{+0.76}_{-0.65}$ & 5\\
        & B   & M0 (2) & 3800$\pm$400 & 1.0$\pm$1.0 & 3.15$\pm$2.00& 130 & 0.76$^{+0.25}_{-0.22}$  &5 \\
HBC 679 & A   & K5 (3) & 4450$\pm$500 & 4.8$\pm$0.7 & 0.00$\pm$0.07& 130 & 0.87$^{+0.29}_{-0.25}$  & 5\\
        & B   & M3 (1) & 3450$\pm$150 & 1.6$\pm$1.0 & 0.00$\pm$0.10& 130 & 0.05$\pm$0.02  & 5 \\
AS 353  & A   & $...$      &   $...$   & $...$   & $...$   &  $...$   & $...$ \\
        & B   & M3 (2) & 3450$\pm$300  & 2.1$\pm$0.8 & 0.02$\pm$0.08 & 150$\pm$50 & 0.68$^{+0.52}_{-0.38}$ & 4\\
\enddata

\tablenotetext{a}{Spectral type to T$_{eff}$ conversions estimated from scale provided in Figure 5 of Luhman (2000).}
\tablenotetext{b}{Distance was not a derived quantity but is included here to show what values were
used in the calculation of $L_*$.  The uncertainty is $\pm$20pc unless otherwise noted.}
\tablenotetext{c}{The approximate uncertainty in the derived spectral type is given in parenthesis, in units of
spectral subclasses.}

\tablerefs{
(1) Perryman et al. 1997;
(2) Krautter et al. 1991;
(3) de Geus et al. 1989;
(4) Edwards \& Snell 1982;
(5) Marraco \& Rydgren 1981.}

\end{deluxetable}

\clearpage

\pagestyle{empty}

\begin{deluxetable}{llrccc}
\tablecolumns{6}
\tablewidth{0pt}
\tablecaption{Component Ages and Masses \label{tbl-8}}
\tablehead{
\colhead{$  $} & \colhead{$  $} & \colhead{Age\tablenotemark{a}} & \colhead{$M$\tablenotemark{b}} & \colhead{$q$} & \colhead{Coeval}\\
\colhead{Name} & \colhead{$  $} & \colhead{(yrs)} & \colhead{(M$_{\odot}$)} & \colhead{($M_2/M_1$)} & \colhead{to 1 $\sigma ?$}}
\startdata
HBC 248 & A & 1$\times$10$^5$ & 2.5 & 0.04 & Y \\
        & B & 2$\times$10$^5$ & 0.1 & $  $ & $  $ \\
HBC 620 & A & 1$\times$10$^6$ & 0.6 & 0.25 & Y\\
        & B & 2$\times$10$^6$ & 0.15 & $  $ & $  $ \\
HBC 625 & A & 3$\times$10$^6$ & 0.4 & 0.38 & Y\\
        & B & 1.5$\times$10$^6$ & 0.15 & $  $ & $  $\\
AS 205  & A & 1$\times$10$^5$ & 1.5 & 0.20 & Y\\
        & B & $<$1$\times$10$^5$ & 0.3 & $  $ & $  $\\
WSB 4   & A & 3$\times$10$^6$ & 0.3 & 1.0 & Y\\
        & B & 3$\times$10$^6$ & 0.3 & $  $ & $  $\\
WSB 19  & A & 1$\times$10$^6$ & 0.3 & 0.5 & Y\\
        & B & 5$\times$10$^5$ & 0.15 & $ $ & $  $ \\
WSB 28  & A & 4$\times$10$^5$ & 0.3 & $\sim$0.2 & Y \\
        & B & $\sim$1$\times$10$^6$\tablenotemark{c} & $\sim$0.06\tablenotemark{c} & $  $ & $  $\\
DoAr 24E& A & 1.5$\times$10$^6$ & $\sim$2.3 & $...$ & $...$\\
        & B & $...$ & $...$ & $  $ & $  $\\
DoAr 26 & A & 4$\times$10$^5$ & 0.15 & 0.67 & Y\\
        & B & 5$\times$10$^5$ & 0.1 & $  $ & $  $\\
ROX 15  & A & $<$1$\times$10$^5$ & 0.3 & 1.0 & Y\\
        & B & 1$\times$10$^5$ & 0.3 & $  $ & $  $\\
SR 21   & A & 1$\times$10$^6$ & 2.5 & 0.06 & N\\
        & B & 1$\times$10$^5$ & 0.15 & $  $ & $  $\\
YLW 15A & A & 1$\times$10$^5$ & 2.5 & $...$ & $...$\\
        & B & $...$ & $...$ & $  $ & $  $\\
H$\alpha$71& A & 5$\times$10$^6$ & 1.5 & 0.07 & N\\
        & B & 2$\times$10$^5$ & 0.1 & $  $ & $  $\\
S CrA   & A & 3$\times$10$^6$ & 1.5 & 0.40 & Y\\
        & B & 1$\times$10$^6$ & 0.6 & $  $ & $  $\\
HBC 679 & A & 7$\times$10$^6$ & 1.2 & 0.25 & Y\\
        & B & 1.5$\times$10$^7$ & 0.3 & $  $ & $  $\\
AS 353  & A & $...$ & $...$ & $...$ & $...$\\
        & B & 5$\times$10$^5$ & 0.3 & $  $ & $  $\\
\enddata

\tablenotetext{a}{Typical uncertainties in age estimates range
from 50$-$200 \%.}
\tablenotetext{b}{Typical uncertainties in mass estimates are
0.1$-$0.3 M$_{\odot}$.}
\tablenotetext{c}{Estimated from position on the evolutionary tracks of
\citet{bar98}, which appear to be roughly consistent with the extrapolation
of the \citet{pal99} tracks for this location.}

\end{deluxetable}

\clearpage

\pagestyle{empty}

\begin{deluxetable}{lllrl}
\tablecolumns{5}
\tablewidth{0pt}
\tablecaption{Circumstellar Disk Diagnostics \label{tbl-9}}
\tablehead{
\colhead{$  $} & \colhead{$  $} & \colhead{$F(Br \gamma)$\tablenotemark{a}} & \colhead{Dereddened $K-L$} & \colhead{$ $} \\
\colhead{Name} & \colhead{$  $} & \colhead{(W m$^{-2}$)} & \colhead{(mag)} & \colhead{Comments}}
\startdata
HBC 248 & A & $...$ &  0.32 $\pm$0.15 &  $  $ \\
        & B & $...$ &  0.48 $\pm$0.18 &  $  $ \\
HBC 620 & A & $...$ &  0.07 $\pm$0.17 & unresolved color\\
        & B & 3.09$\times$10$^{-18} \pm$2.00$\times$10$^{-18}$ & $  $ & chromospheric ?\\
HBC 625 & A & $...$ &  0.07 $\pm$0.09 & unresolved color\\
        & B & $...$ &  $  $ & $  $\\
AS 205  & A & 7.19$\times$10$^{-16} \pm$2.43$\times$10$^{-16}$ & 0.83 $\pm$0.16 &$  $\\
        & B & 1.53$\times$10$^{-16} \pm$0.90$\times$10$^{-16}$ & 0.98 $\pm$0.19 &$  $\\
WSB 4   & A & $...$ & 0.44 $\pm$0.18 &  $  $\\
        & B & 2.87$\times$10$^{-18} \pm$2.87$\times$10$^{-18}$ & 0.96 $\pm$0.18 & $  $\\
DoAr 24E& A & $...$ & 0.40 $\pm$0.19 & $  $\\
        & B & 2.26$\times$10$^{-17} \pm$2.41$\times$10$^{-17}$ & 2.2 $\pm$0.1&A$_V$ unknown\\
ROX 15  & A & $...$ & 0.13 $\pm$0.19 & unresolved color\\
        & B & $...$ & $  $ & $  $\\
SR 21   & A & absorption & 0.64 $\pm$0.12 & $  $\\
        & B & $...$ & $-$0.09 $\pm$0.14 & $  $\\
YLW 15  & A & 1.56$\times$10$^{-15} \pm$1.20$\times$10$^{-15}$ & 0.86 $\pm$0.12 & A$_V=$38 mag (see text)\\
        & B & $...$ & 1.69 $\pm$0.08 & A$_V$ unknown\\
WSB 71& A & $...$ & 1.14 $\pm$0.21 & $  $\\
        & B & 1.23$\times$10$^{-17} \pm$0.68$\times$10$^{-17}$ & 0.37 $\pm$0.24 & chromospheric ?\\
S CrA   & A & 4.23$\times$10$^{-16} \pm$1.10$\times$10$^{-16}$ & 1.44 $\pm$0.13 &$  $\\
        & B & 1.59$\times$10$^{-16} \pm$0.57$\times$10$^{-16}$ & 1.14 $\pm$0.15 &$  $\\
HBC 679 & A & $...$ & $-$0.33 $\pm$0.14 & unresolved color\\
        & B & $...$ & $  $ & $  $\\
VV CrA (1996)& A &8.07$\times$10$^{-16} \pm$0.52$\times$10$^{-16}$&0.99 $\pm$0.06 & A$_V$ unknown\\
        & B & 1.70$\times$10$^{-16} \pm$0.49$\times$10$^{-16}$ & 3.63 $\pm$0.06 & A$_V$ unknown\\
VV CrA (1997)& A & 7.33$\times$10$^{-16} \pm$0.52$\times$10$^{-16}$ & 0.99 $\pm$0.06 &A$_V$ unknown\\
        & B & 1.93$\times$10$^{-16} \pm$0.49$\times$10$^{-16}$ & 3.63 $\pm$0.06 & A$_V$ unknown\\
AS 353  & A & 4.51$\times$10$^{-16} \pm$0.21$\times$10$^{-16}$ & 1.88 $\pm$0.03 &A$_V$ unknown\\
        & B & $...$ & 0.43 $\pm$0.13 & $  $ \\
\enddata

\tablenotetext{a}{The uncertainty in the Br $\gamma$ emission line flux
is based on a 1 \AA~ uncertainty in the measured equivalent width.}

\end{deluxetable}

\clearpage

\newpage
\figcaption[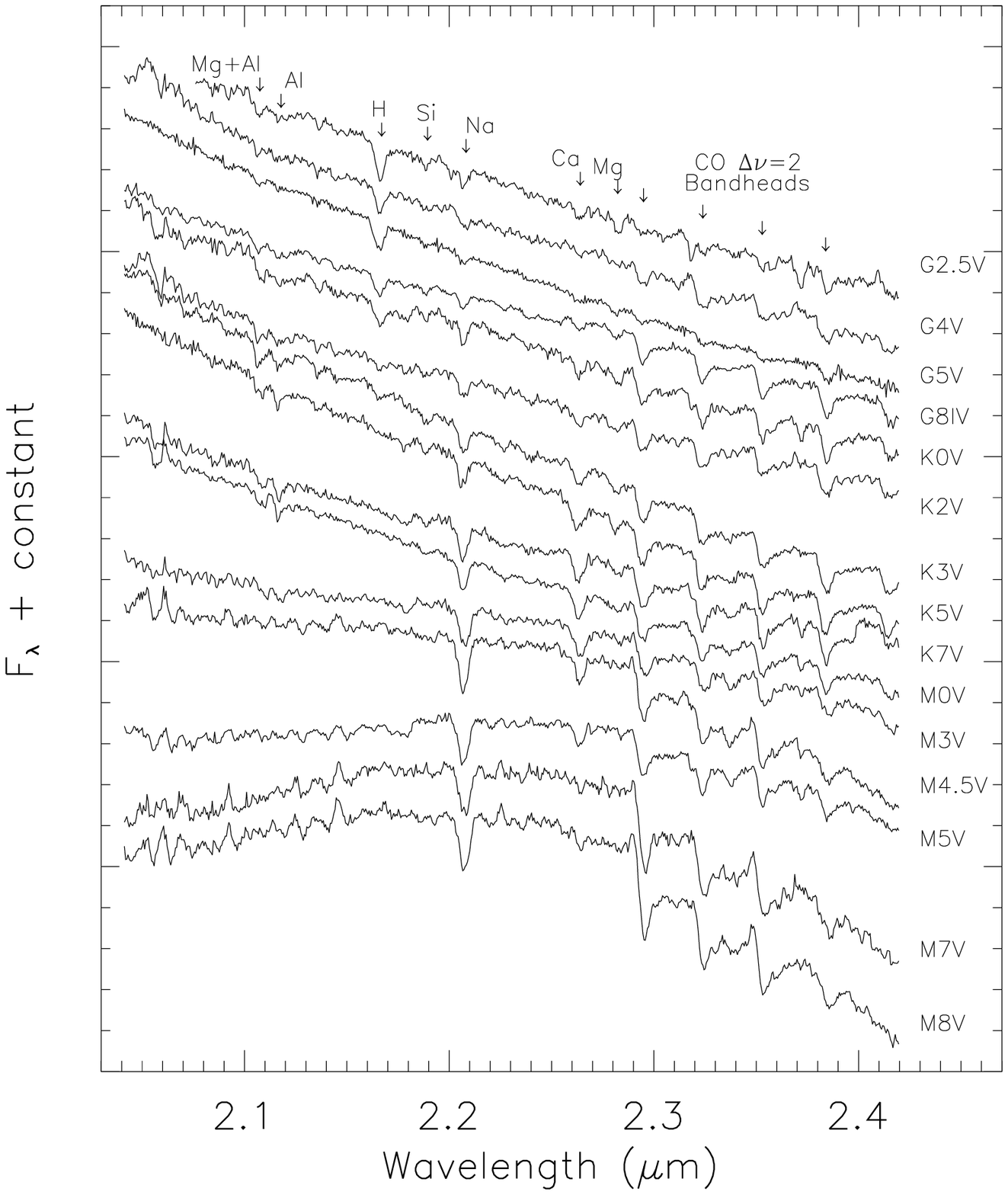]{$K$-band spectra of main-sequence spectral type standards.
The resolution is 760.  The stars are identified in Table 2.
Major spectral features are indicated; details of these transitions
appear in Table 3.}

\figcaption[f2.eps]{$K$-band spectra of sample objects. 
The resolution is 760.  The stars are identified in Table 1.
Major spectral features are indicated at the top of each
column of spectra; the positions of the He I and H$_2$ lines are indicated on
the 1996 AS 353 and VV CrA spectra.  For each system, the primary star
spectrum is plotted above the secondary.  All spectra were normalized to
unity at 2.2 $\mu$m
and offset by an additive constant for presentation purposes.
The scaling does not necessarily reflect true flux ratios.}

\figcaption[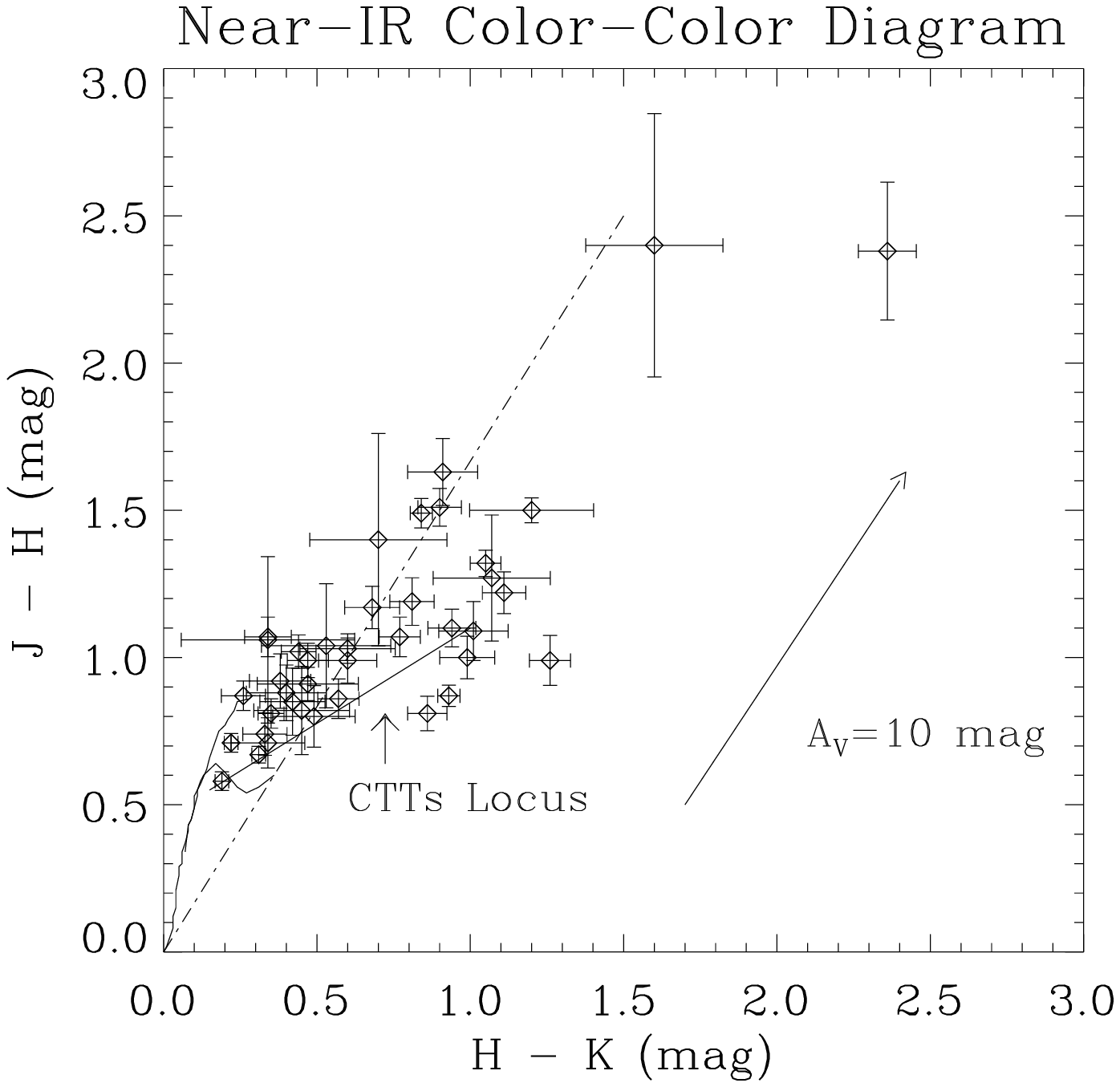]{The $J-H$, $H-K$ color-color
diagram of the young star sample, excepting YLW 15A.
The dash-dot line delineates between objects with (right hand
side) and without a near-IR excess above photospheric values for colors, 
attributable to circumstellar material.
The dwarf and giant star loci are overplotted.  An A$_V=10$ reddening
vector and
the CTTs locus of Meyer et al. (1997) and are indicated; to the
left of the dash-dot line, this locus is representative of WTTS,
which have main-sequence dwarf colors.}

\figcaption[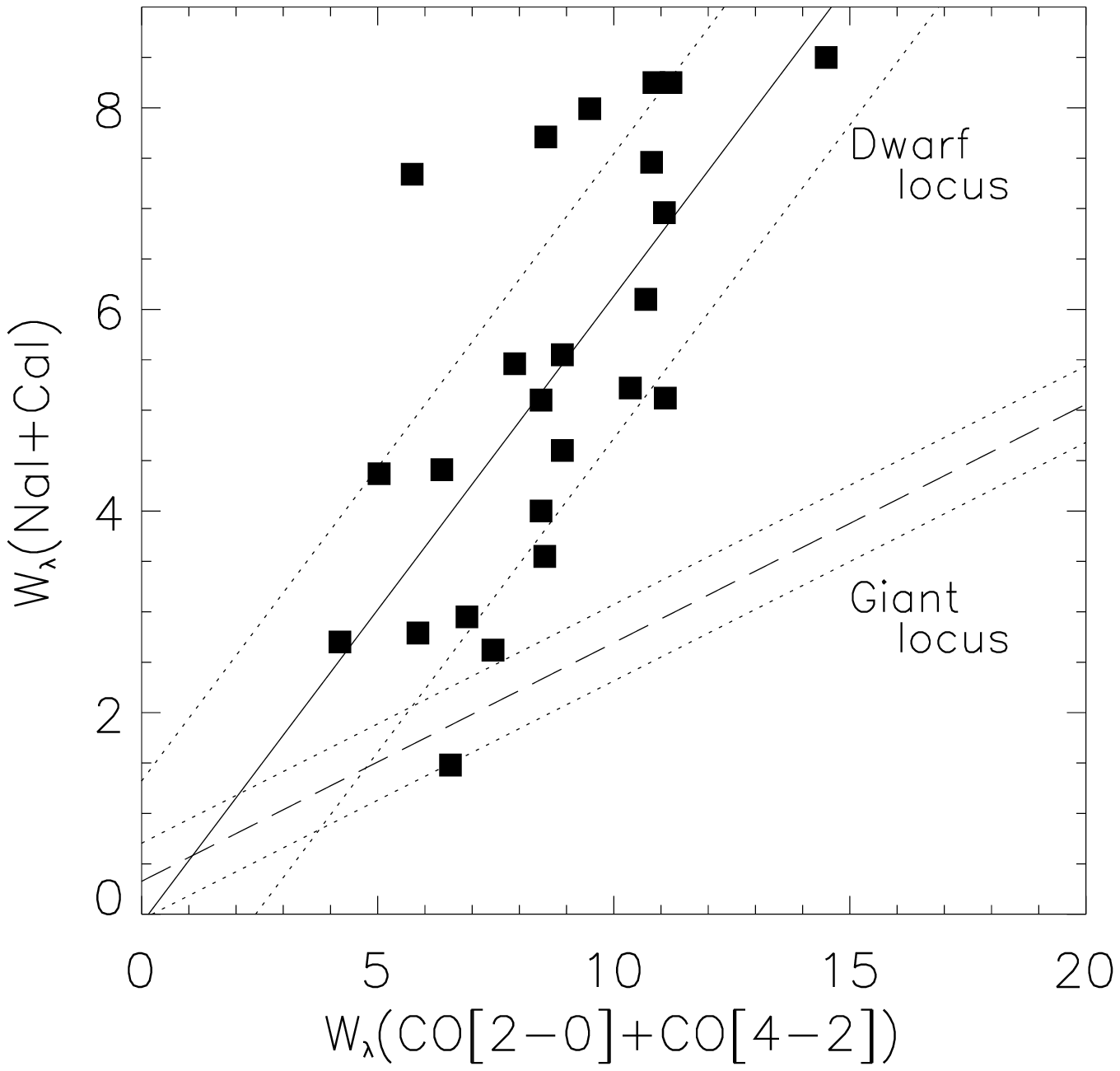]{The sum
of the Na I $+$ Ca I {\it versus} the CO(2$-$0) $+$ CO(4$-$2)
equivalent widths.  The solid line shows the main-sequence (dwarf)
locus, and the dashed line the giant locus.  These loci are fits to
the dwarf data, given in Table 5, and to the giant star equivalent widths
calculated from the smoothed spectra of Wallace \& Hinkle (1997).
1 $\sigma$ errors for
the fit of the loci to the stellar data are shown as dotted lines.
The PMS binary components are plotted as solid squares.}

\figcaption[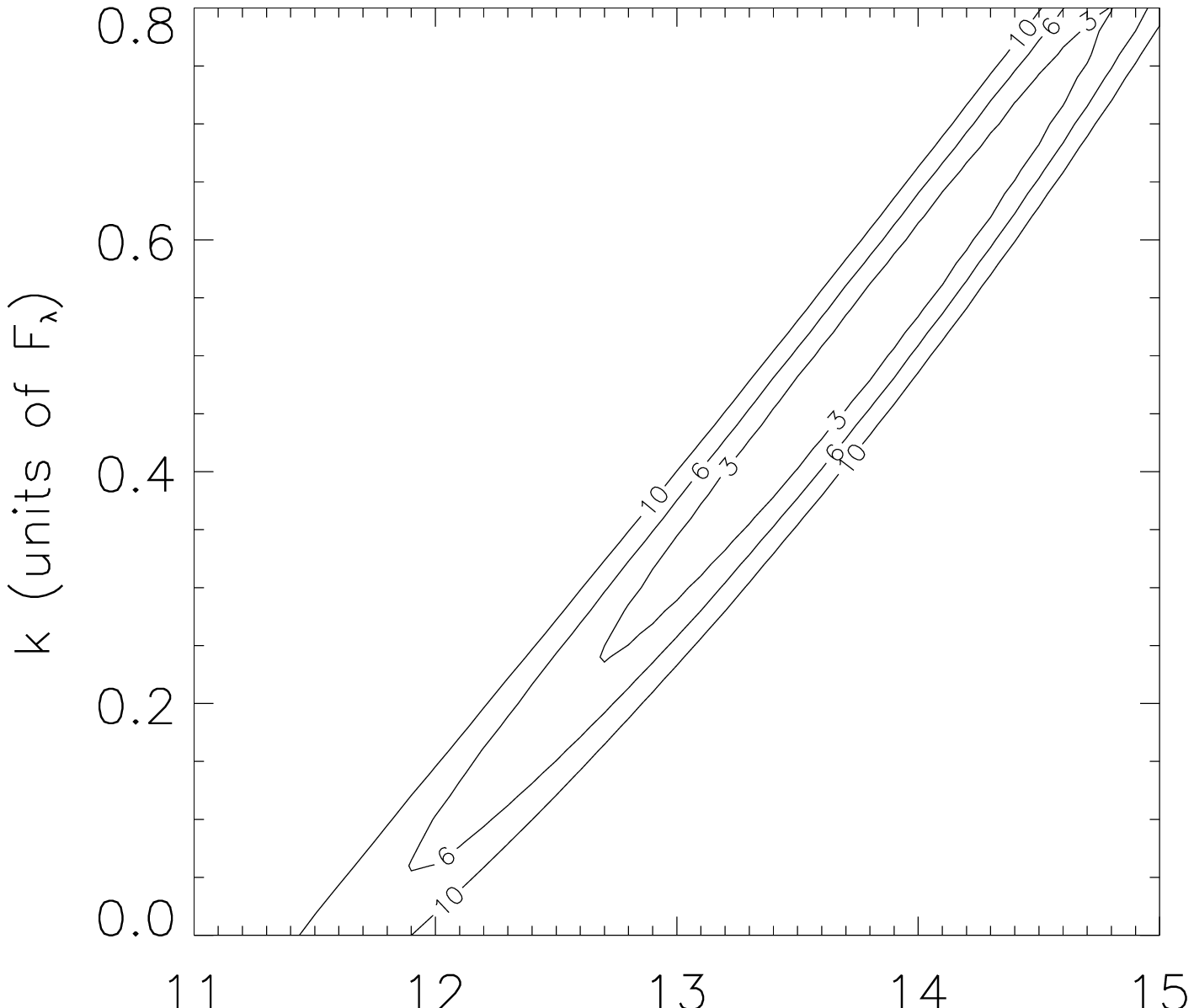]{Contour plot of reduced $\chi^2$ for
a range of values of $A_v$ and $k$ calculated for the fit of
the ROX 15 secondary to the M3 standard, for a given value of the
constant, $c$.}

\figcaption[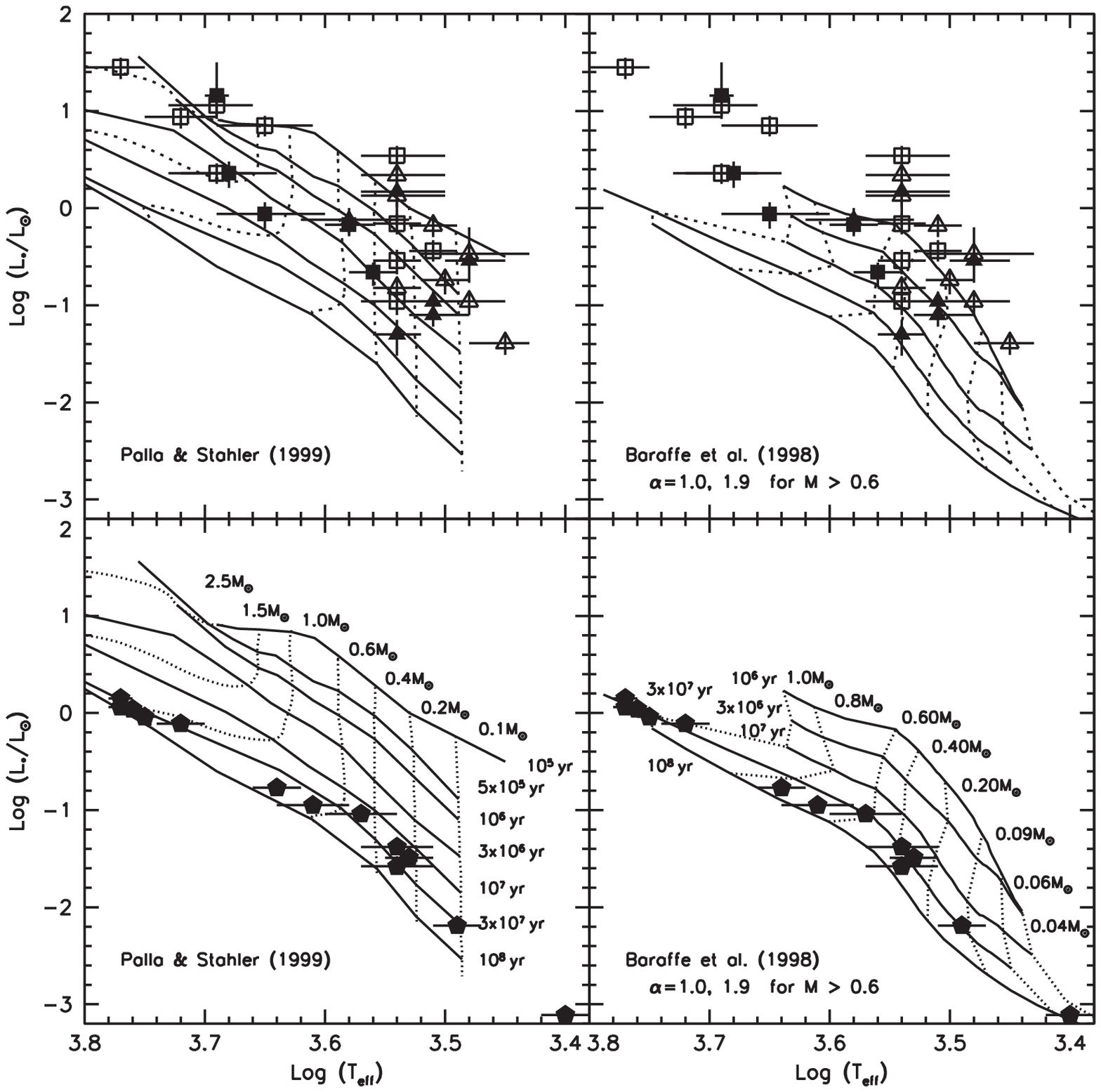]{$T_{eff}$ and luminosity for the young star
binary components plotted on the PMS evolutionary
tracks of Palla \& Stahler (1999) and Baraffe et al. (1998).
The young star sample is plotted in the upper panels; the open
symbols represent objects in the Ophiuchus SFR.  Primary stars
are plotted as squares and secondaries as triangles.
A sample of main-sequence spectral type standard stars appears
in the lower panels.  The mass tracks and isochrones are
labelled in the lower panels.
For objects lying between labelled isochrones
and mass tracks, the ages and masses were estimated
by interpolation.}

\figcaption[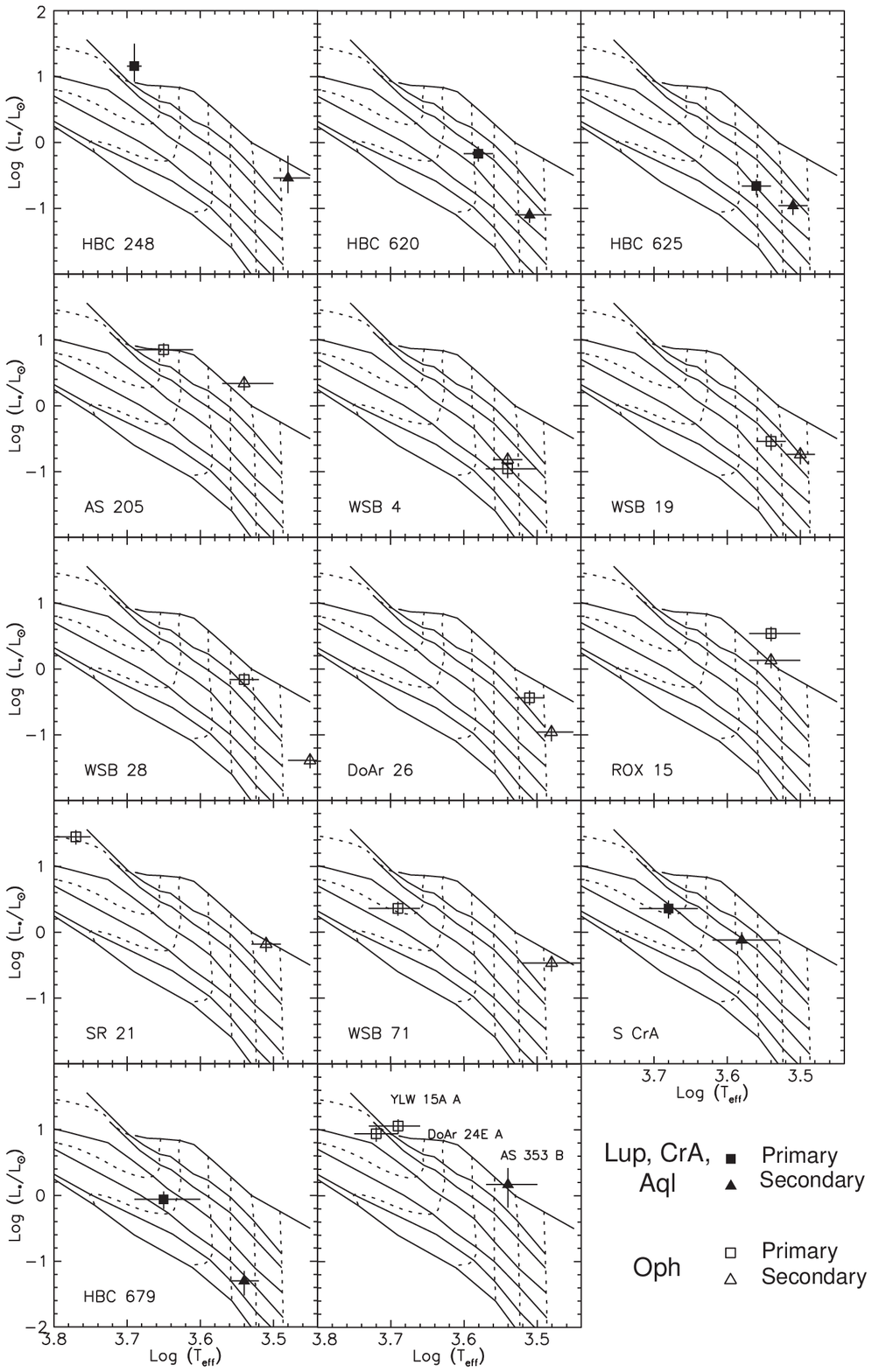]{Same
as the upper, left hand panel in Figure 6 except
each binary pair is plotted individually, as labelled.  The
mass tracks and isochrones are the same as labelled in
Figure 6, lower left hand panel.}

\figcaption[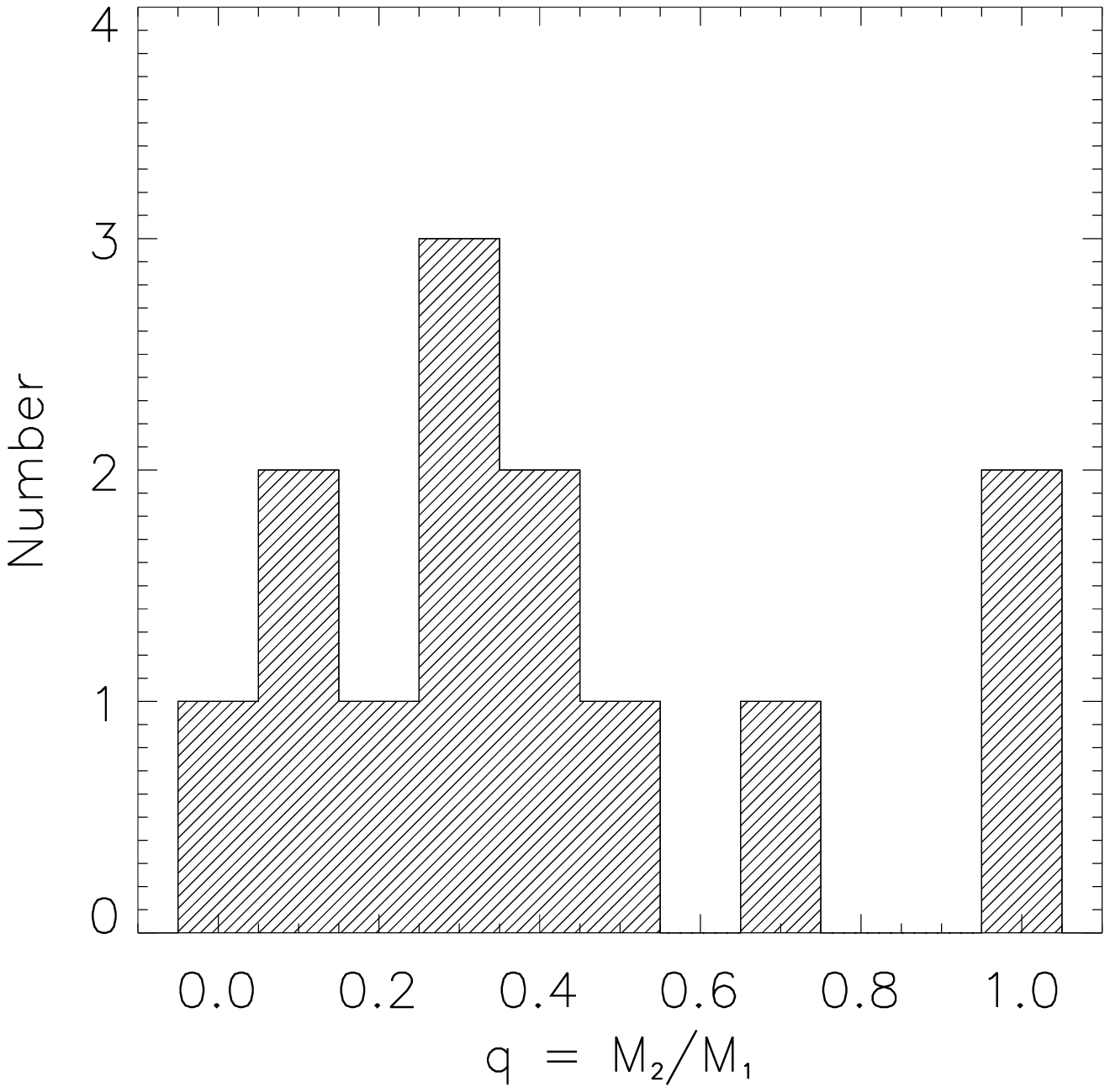]{Histogram of M$_2$/M$_1$ for the
13 systems with measured masses for both components.  The $q=0$
bin contains one object with $q<0.05$.}

\figcaption[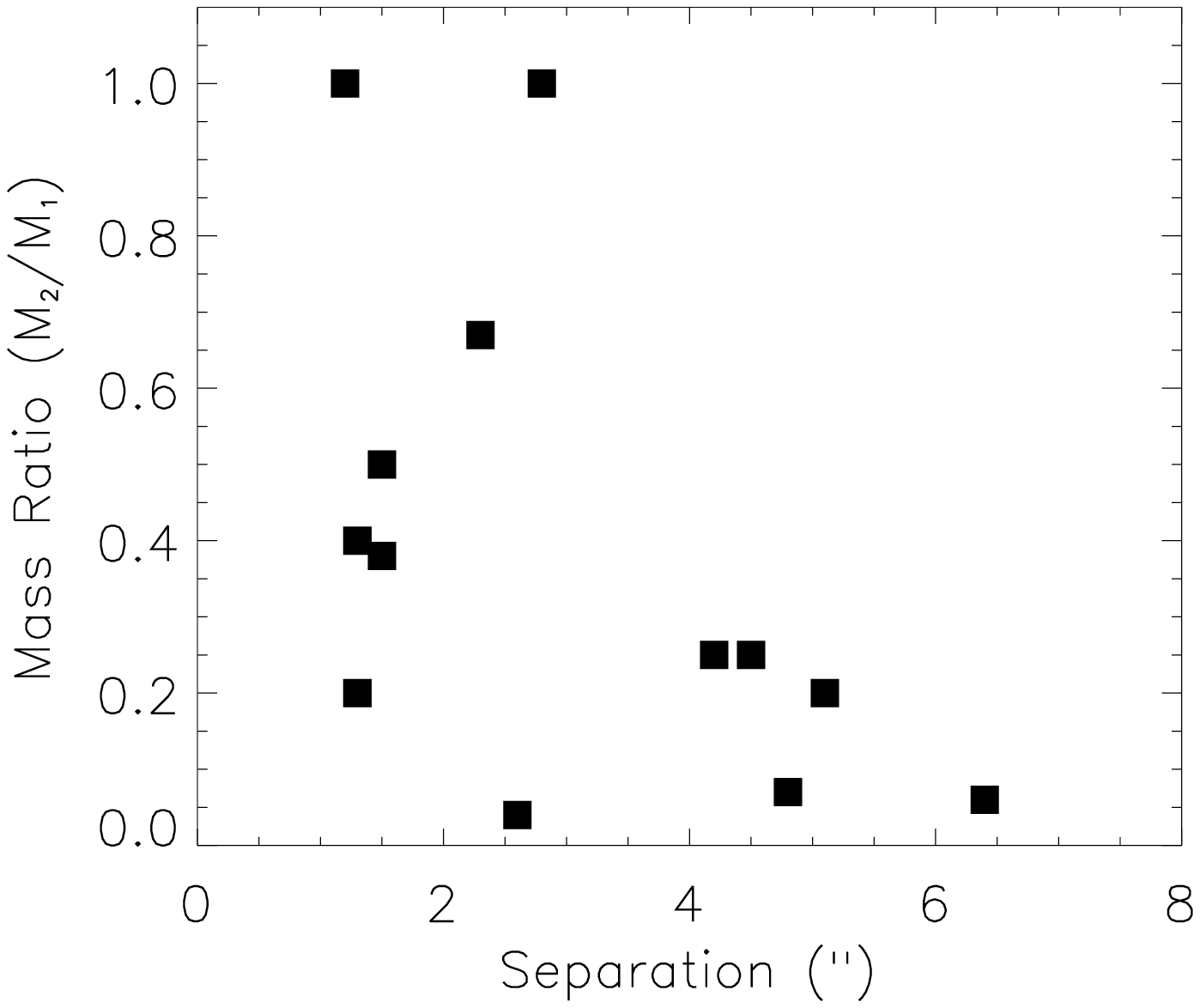]{Mass ratio as a function of binary separation.
Uncertainties in the mass ratios are not shown but are as high as
$\sim$100 \% in a few cases.  Nonetheless, a trend towards
lower mass ratios at larger separations appears to be present.}

\figcaption[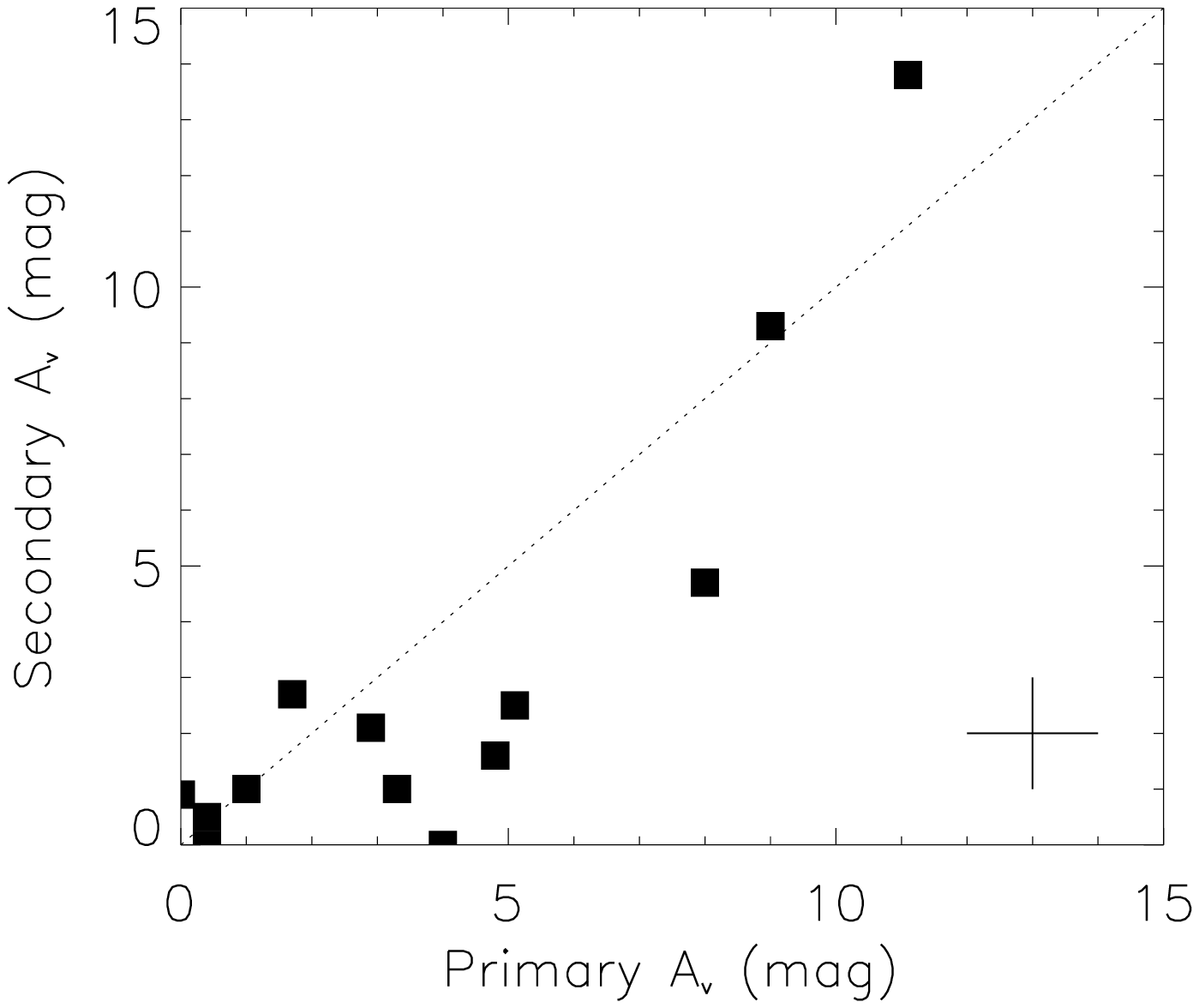]{The primary {\it versus} secondary $A_v$'s
for all the stars in the spectroscopic sample except the
components in DoAr 24 E, YLW 15 A, VV CrA, and AS 353.  A typical $\pm$1 mag
error bar is shown in the lower right hand corner.  The dotted line
delineates equal component extinctions.}

\figcaption[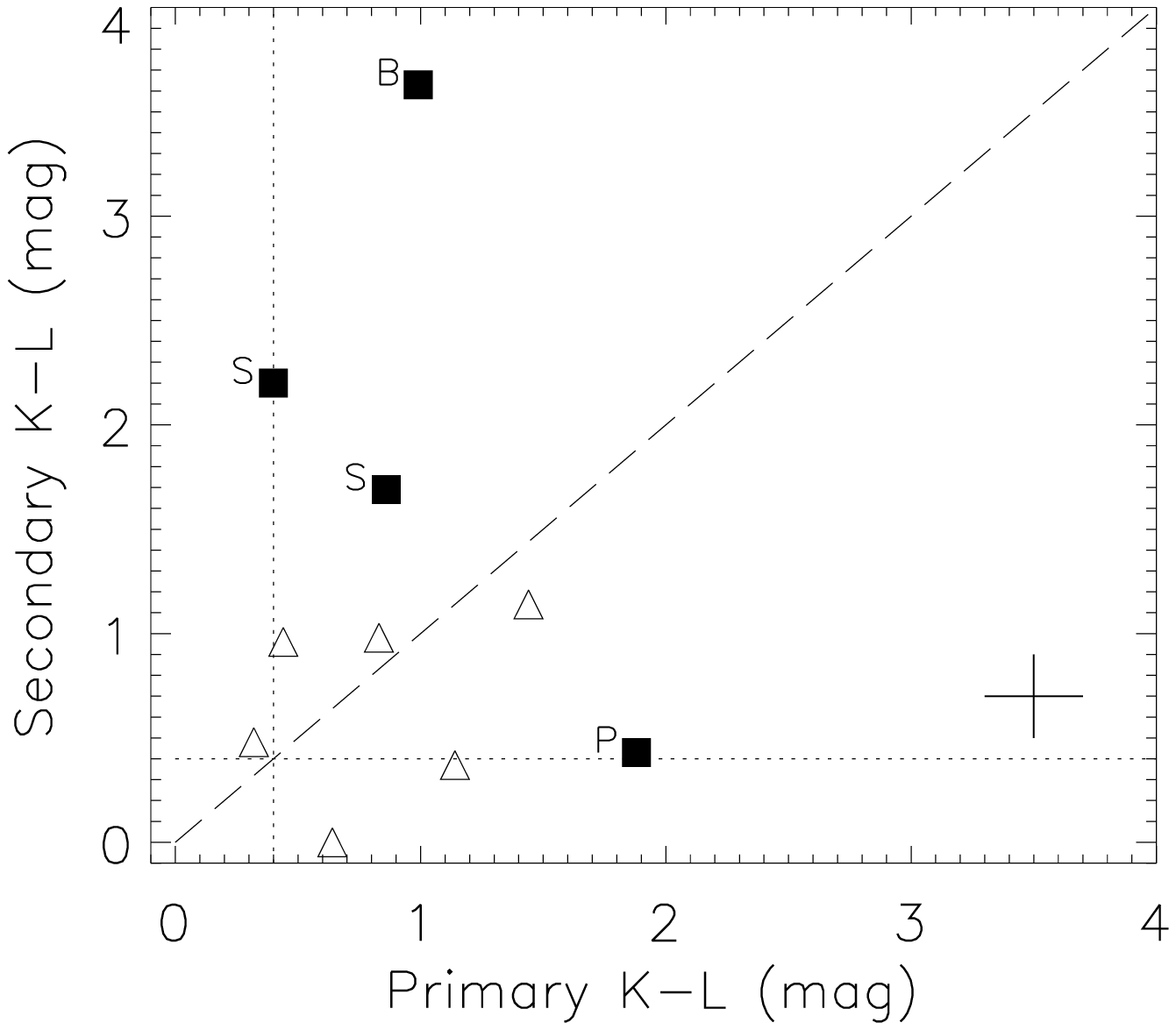]{The primary {\it versus} secondary $K-L$
colors as available for the sample objects
(Table 9).  {\it Open triangles:} systems for which both
components have been dereddened.  {\it Filled squares:} Systems for
which it was not possible to deredden both components; the letters
next to each symbol indicate which component was {\it not} dereddened:
primary (P), secondary (S), or both (B).  A typical uncertainty is
shown in the lower right hand corner.  The dashed line
delineates equal component $K-L$ colors.  The dotted line shows
the upper limit of the CTT/WTT 0.4 mag cutoff of \citet{edw93}
(see also \S 4.3.1).}

\figcaption[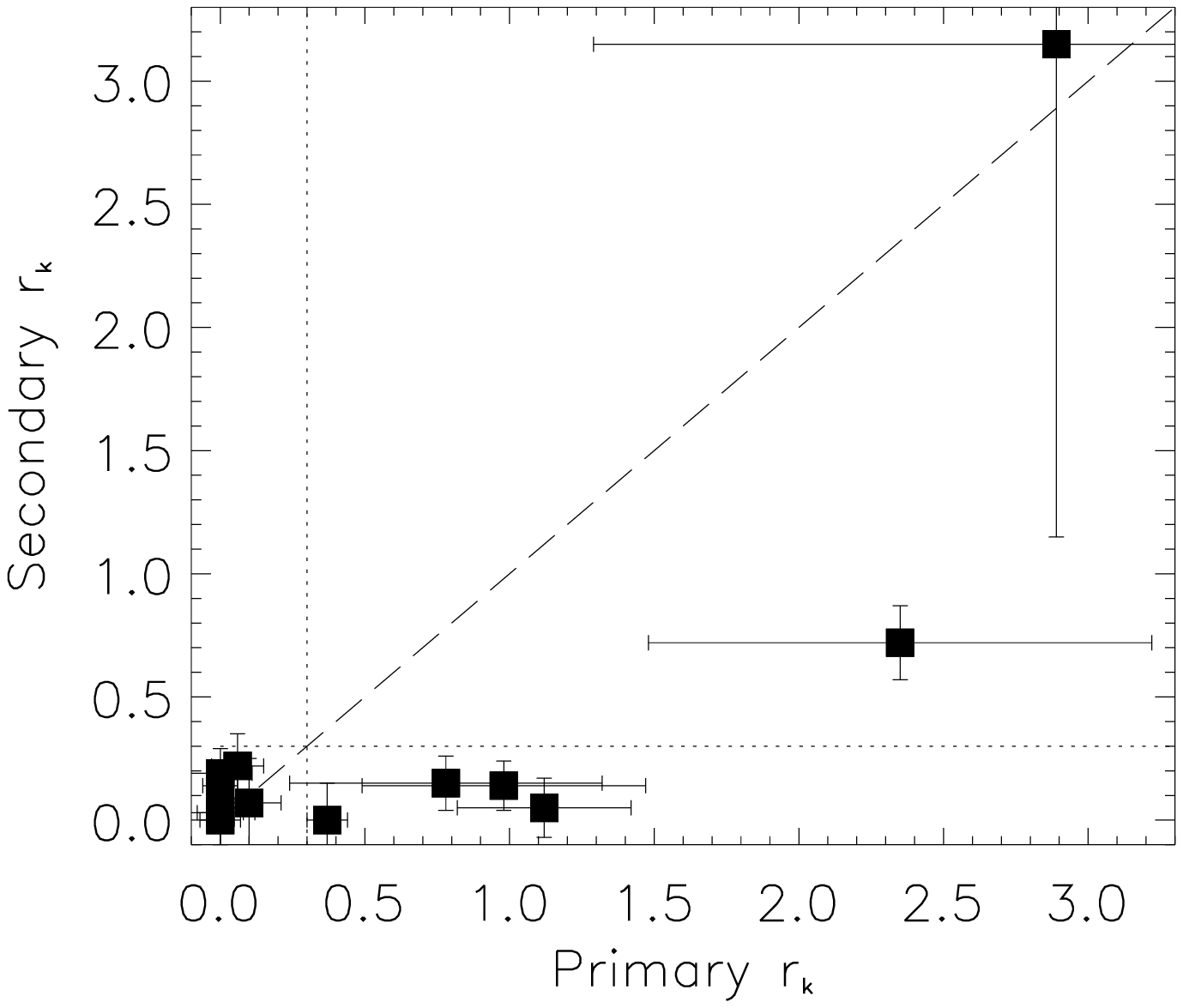]{The primary {\it versus} secondary $r_k$
veiling parameter for the objects as described in the Figure 10 caption.
The dashed line delineates equal component $r_k$ values.  The dottted
lines indicate probable cutoff between objects with and without
a $K$-band excess.}

\figcaption[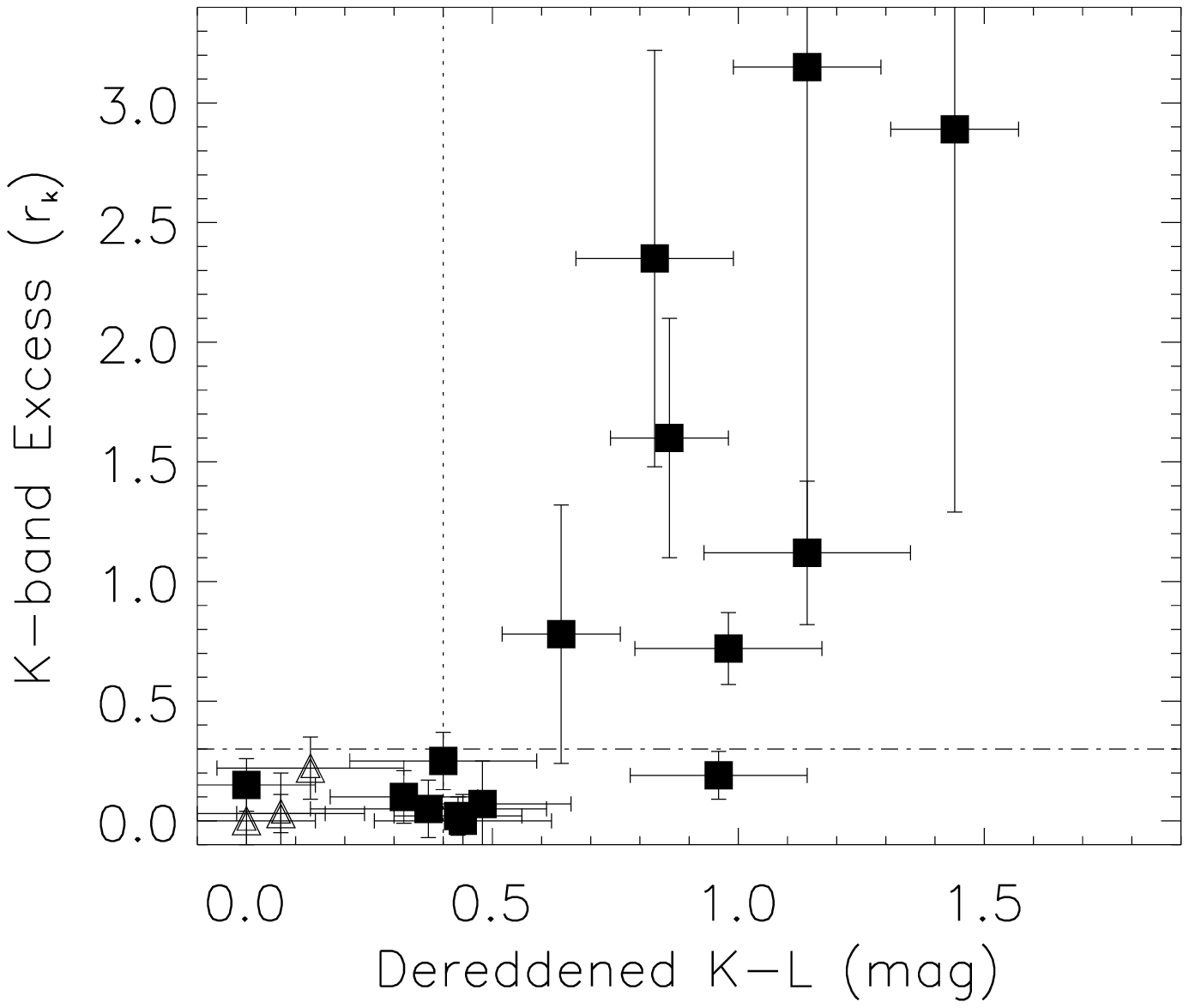]{The $K$-band excess, $r_k$, plotted as a function
of the $K-L$ color of the star.  In several cases, where component
resolved $K-L$ colors of the systems were not available, we
plot the $K-L$ total for the system and the higher of the
$r_k$ values for the two stars (Table 7).  These are designated
with an open triangle.  The dot-dash and dotted lines represent
the cutoffs in the respective quantities for CTT {\it versus} WTT
behavior.}

\figcaption[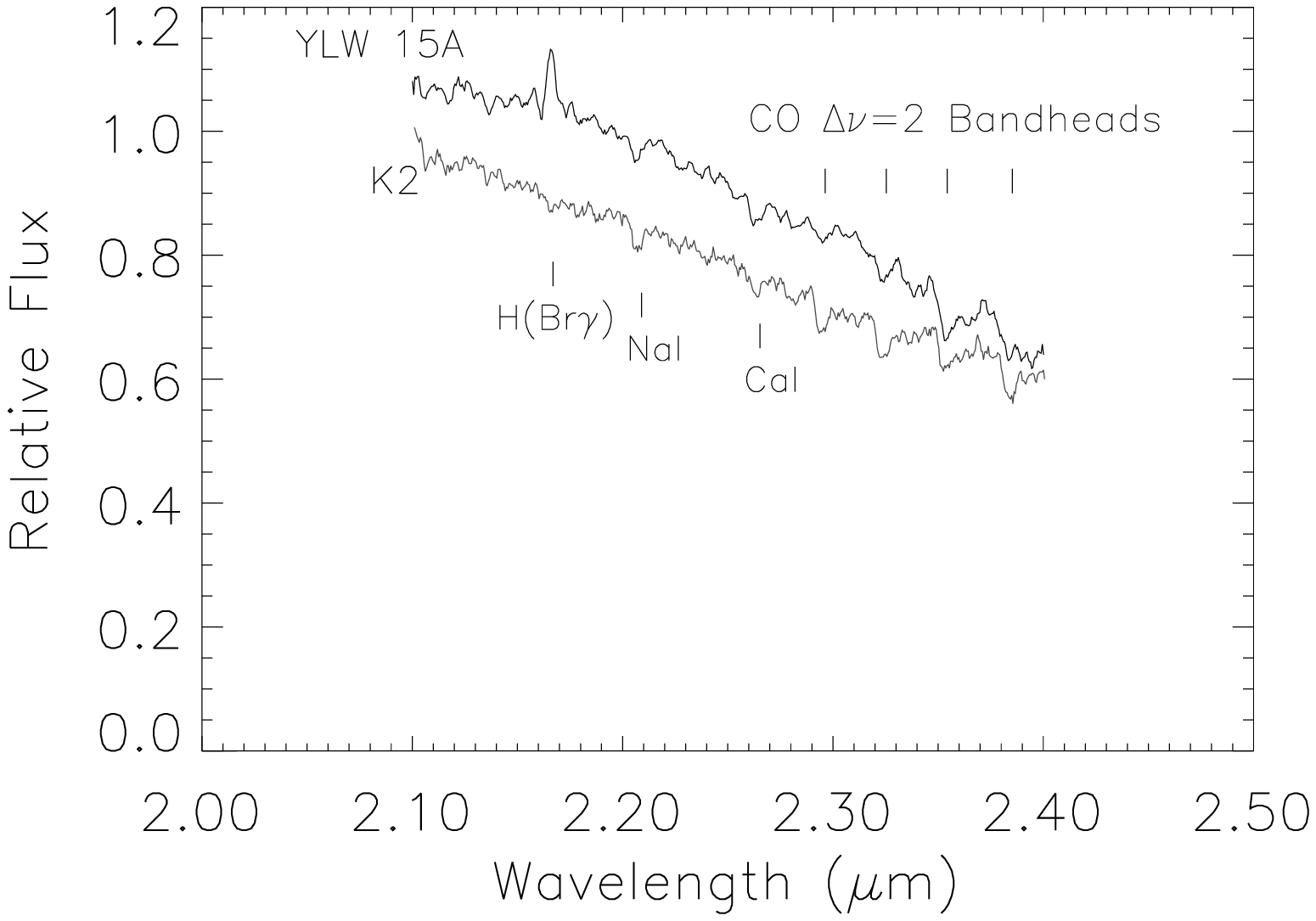]{The spectrum of YLW 15A, unextinguished by $A_V=$38,
with a $K$-band excess of 1.7 subtracted, and smoothed by a boxcar of
width 5 pixels ($\sim$1 resolution element).  The lower spectrum is that
of a K2 standard star.  The observed spectrum of YLW 15A appears in 
Figure 2.}

\clearpage

\begin{figure}
\figurenum{1}
\plotone{f1.eps}
\end{figure}

\begin{figure}
\figurenum{1}
\plotone{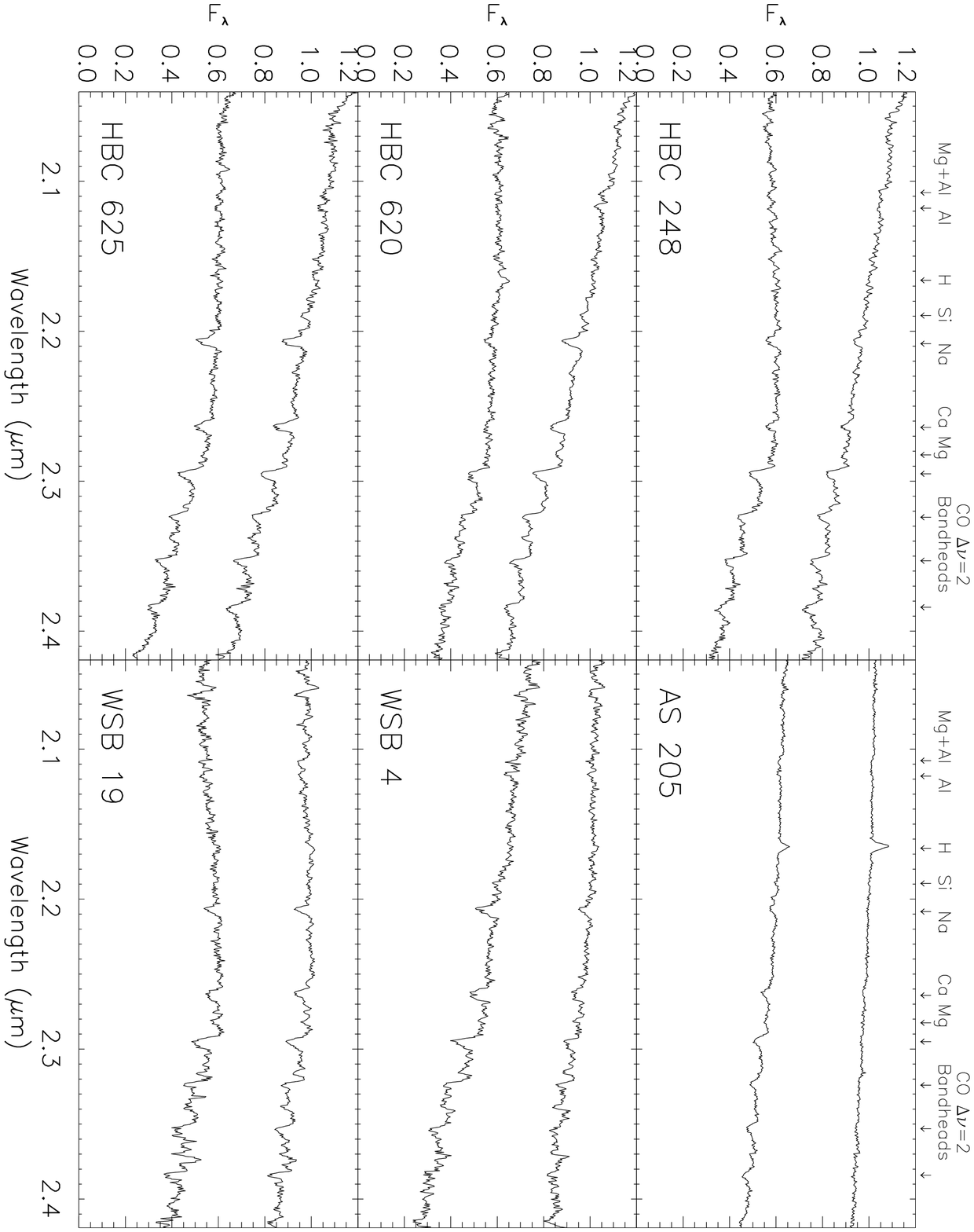}
\end{figure}

\begin{figure}
\figurenum{1}
\plotone{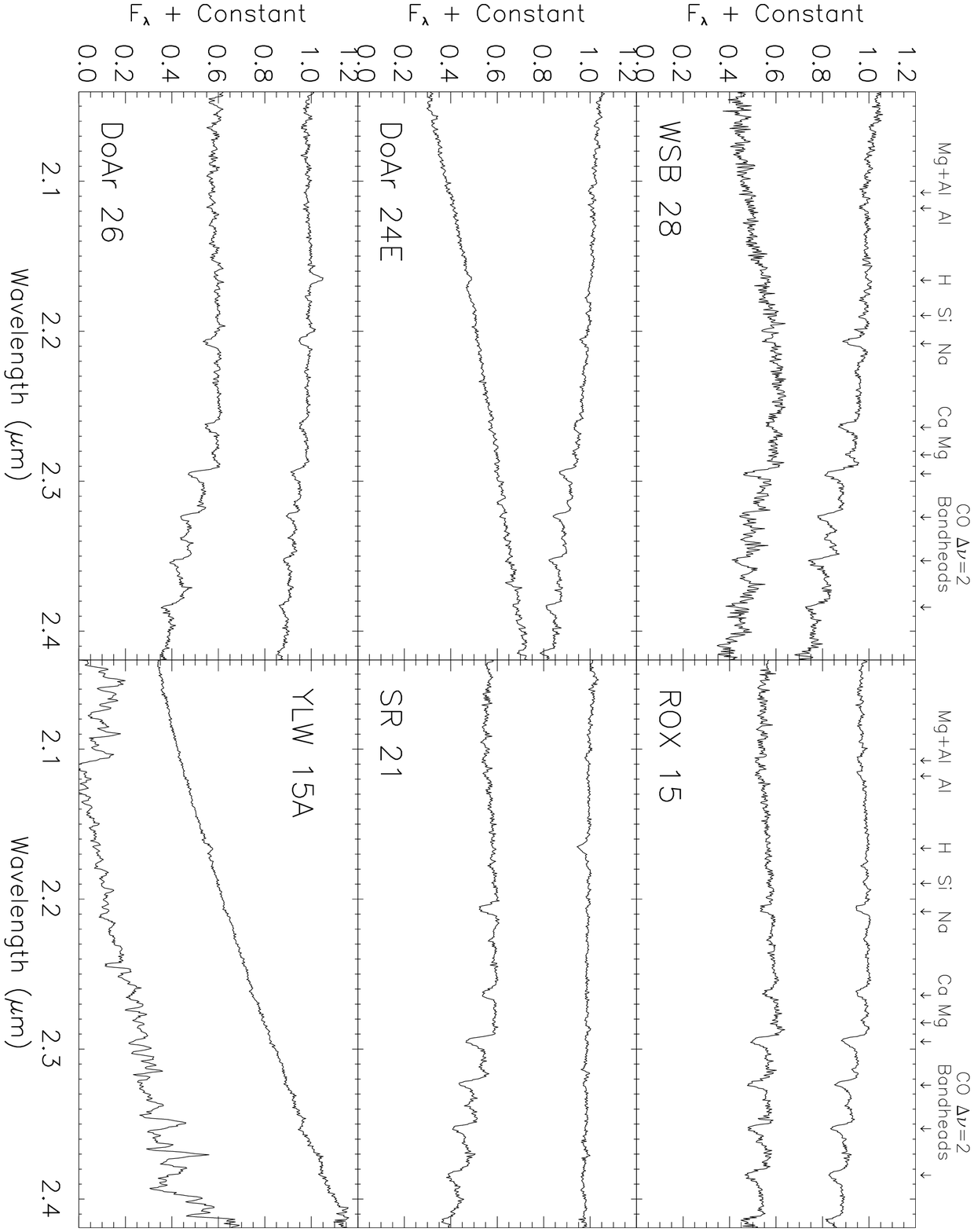}
\end{figure}

\begin{figure}
\figurenum{1}
\plotone{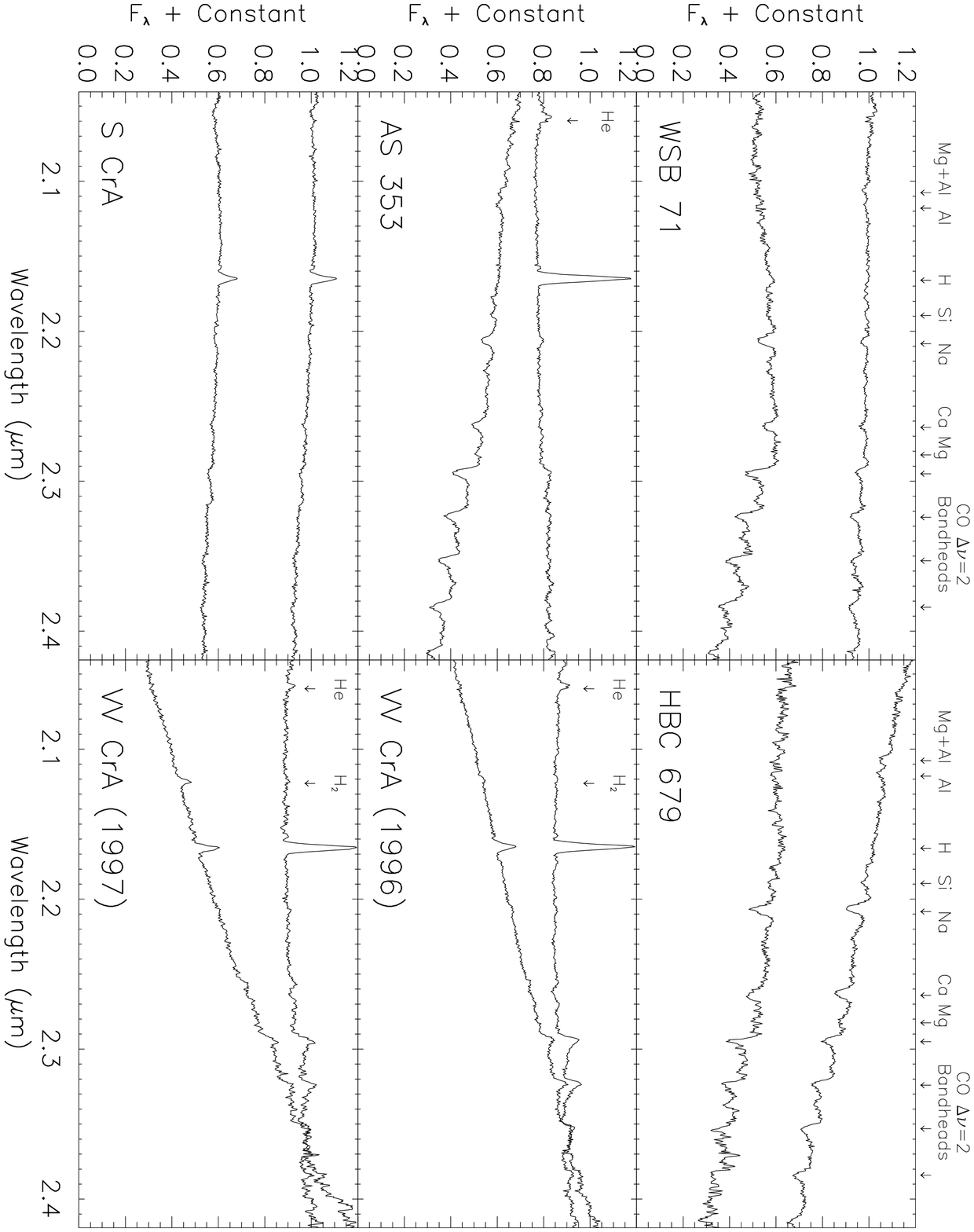}
\end{figure}

\begin{figure}
\figurenum{1}
\plotone{f3.eps}
\end{figure}

\begin{figure}
\figurenum{1}
\plotone{f4.eps}
\end{figure}

\begin{figure}
\figurenum{1}
\plotone{f5.eps}
\end{figure}

\begin{figure}
\figurenum{1}
\plotone{f6.eps}
\end{figure}

\begin{figure}
\figurenum{1}
\plotone{f7.eps}
\end{figure}

\begin{figure}
\figurenum{1}
\plotone{f8.eps}
\end{figure}

\begin{figure}
\figurenum{1}
\plotone{f9.eps}
\end{figure}

\begin{figure}
\figurenum{1}
\plotone{f10.eps}
\end{figure}

\begin{figure}
\figurenum{1}
\plotone{f11.eps}
\end{figure}

\begin{figure}
\figurenum{1}
\plotone{f12.eps}
\end{figure}

\begin{figure}
\figurenum{1}
\plotone{f13.eps}
\end{figure}

\begin{figure}
\figurenum{1}
\plotone{f14.eps}
\end{figure}

\end{document}